\documentclass[%
reprint,
superscriptaddress,
showpacs,preprintnumbers,
amsmath,amssymb,
prb,
]{revtex4-2}
\usepackage{graphicx}
\usepackage{dcolumn}
\usepackage{xfrac}
\usepackage{bm}
\usepackage{color}
\usepackage[export]{adjustbox}
\usepackage{xr-hyper}
\usepackage{hyperref}
\usepackage{tabularx}
\usepackage{easyReview}

\begin{document}
	
	
	\title{Absence of superconductivity in electron-doped chromium pnictides ThCrAsN$_{1-x}$O$_x$} 
	
	
	\author{Zhi-Cheng Wang}
	\email{wzc@seu.edu.cn}
	\affiliation{Key Laboratory of Quantum Materials and Devices of Ministry of Education, School of Physics, Southeast University, Nanjing 211189, China}
	\author{Ye-Ting Shao}
	\affiliation{School of Physics, Interdisciplinary Center for Quantum Information and State Key Laboratory of Silicon and Advanced Semiconductor Materials, Zhejiang University, Hangzhou 310058, China}
	\author{Yi-Qiang Lin}
	\affiliation{School of Physics, Interdisciplinary Center for Quantum Information and State Key Laboratory of Silicon and Advanced Semiconductor Materials, Zhejiang University, Hangzhou 310058, China}
	\author{Shi-Jie Song}
	\affiliation{School of Physics, Interdisciplinary Center for Quantum Information and State Key Laboratory of Silicon and Advanced Semiconductor Materials, Zhejiang University, Hangzhou 310058, China}
	\author{Bai-Zhuo Li}
	\affiliation{School of Physics, Interdisciplinary Center for Quantum Information and State Key Laboratory of Silicon and Advanced Semiconductor Materials, Zhejiang University, Hangzhou 310058, China}
	\author{Er-Jian Cheng}
	\affiliation{State Key Laboratory of Surface Physics and Department of Physics, Fudan University, Shanghai 200438, China}
	\author{Shi-Yan Li}
	\affiliation{State Key Laboratory of Surface Physics and Department of Physics, Fudan University, Shanghai 200438, China}
	\affiliation{Collaborative Innovation Centre of Advanced Microstructures, Nanjing University, Nanjing 210093, China}
	\affiliation{Shanghai Research Center for Quantum Sciences, Shanghai 201315, China}
	\author{Qin-Qing Zhu}
	\affiliation{School of Science, Westlake University, Hangzhou 310024, China}
	\affiliation{Institute of Natural Sciences, Westlake Institute for Advanced Study, Hangzhou 310024, China}
	\author{Zhi Ren}
	\affiliation{School of Science, Westlake University, Hangzhou 310024, China}
	\affiliation{Institute of Natural Sciences, Westlake Institute for Advanced Study, Hangzhou 310024, China}
	\author{Guang-Han Cao}	
	\email{ghcao@zju.edu.cn}
	\affiliation{School of Physics, Interdisciplinary Center for Quantum Information and State Key Laboratory of Silicon and Advanced Semiconductor Materials, Zhejiang University, Hangzhou 310058, China}
	\affiliation{Collaborative Innovation Centre of Advanced Microstructures, Nanjing University, Nanjing 210093, China}

	
	\date{\today}
	
	\begin{abstract}
		 Theoretical studies predicted possible superconductivity in electron-doped chromium pnictides isostructural to their iron counterparts. Here, we report the synthesis and characterization of a new ZrCuSiAs-type Cr-based compound ThCrAsN, as well as its oxygen-doped variants. All samples of ThCrAsN$_{1-x}$O$_x$ show metallic conduction, but no superconductivity is observed above 30 mK even though the oxygen substitution reaches 75\%. The magnetic structure of ThCrAsN is determined to be G-type antiferromagnetic by magnetization measurements and first-principles calculations jointly. The calculations also indicate that the in-plane Cr--Cr direct interaction of ThCrAsN is robust against the heavy electron doping. The calculated density of states of the orbital occupations of Cr for ThCrAs(N,O) is strongly spin-polarized. Our results suggest the similarities between chromium pnictides and iron-based superconductors shouldn't be overestimated.
	
	\end{abstract}
	
	
	\maketitle
	
	\section{Introduction}
	Since the discovery of high-$T_c$ superconductivity in Fe-based materials~\cite{LaFeAsO_2008}, tremendous efforts have been devoted to search for superconductivity in the Fe-free transition-metal systems with the conducting layers isostructural to Fe$_2X_2$ ($X$ = As or Se) motifs~\cite{Review_2011,Review_2019}. Among them Cr$_2$As$_2$-layer-based compounds are of particular interest because they exhibit antiferromagnetism with high Neel temperature ~\cite{LaCrAsO_2013,BaCr2As2_2009,EuCr2As2_2014,SrCr2As2_2017,Sr2Cr3As3O2,Sr2ScCrAsO3,BaTiCrAsO,Sr2Cr2AsO3}, even though Cr-based superconductors are scarce due to the robustness of antiferromagnetism~\cite{CrReview_2019}. The exploration of superconductivity in Cr$_2$As$_2$-layer-based materials is not only inspired by the breakthrough in Cr-based superconductors in recent years~\cite{CrAs_2014,K2Cr3As3_2015,KCr3As3_2017,K1331_2019,Pr3Cr10-xN11_2020}, but also supported by the theoretical predictions~\cite{BaCr2As2_2017,LaCrAsO_2017_Pizarro,LaCrAsO_2017_Wang}. 
	
	Within the framework of the Mott scenario for the transition-metal arsenides~\cite{MottScenario_2011}, the $d^4$ case of Cr$^{2+}$ is symmetrical to the $d^6$ case of Fe$^{2+}$ with respect to half-filled $d^5$ configuration (Mn$^{2+}$). In this context, Fe-based superconductors (FeSCs) are electron-doped systems compared to the Mott-type parent materials. Accordingly, Cr-based compounds are expected to show comparable electronic correlations with possible superconductivity as the hole-doped side to the $d^5$ system, when the $3d^n$ fillings is between $n=4$ and $n=5$~\cite{LaCrAsO_2017_Pizarro,LaCrAsO_2017_Wang}. That is to say, electron doping in the $d^4$ configuration is likely to bring about superconductivity in Cr$_2$As$_2$-based materials.
	
	However, it's not easy to find a proper Cr$_2$As$_2$-based compound to study the electron-doping effect systematically. With regard to LaCrAsO, the solid solubility limits F$^-$ substitution for O$^{2-}$ to 20\%~\cite{LaCrAsO_2013}, while H$^-$ doping results in the structural transformation~\cite{La2Cr2As2OyHx_2017}. In addition, the replacements of Cr site by Mn and Fe lead to a metal--insulator transition in LaCrAsO and BaCr$_2$As$_2$, respectively~\cite{LaCrAsO_2013,BaCrFeAs_2017}. 
	
	In this work, we report a new chromium pnictide, ThCrAsN, which is isostructural to LaCrAsO. The advantage of Th$_2$N$_2$ layers over La$_2$O$_2$ layers is the high solubility of O$^{2-}$ in N$^{3-}$, making high electron doping possible~\cite{ThFeAsN,ThFeAsN1-xOx,BaTh}. A series of polycrystalline samples of ThCrAsN$_{1-x}$O$_x$ were synthesized, and the actual O$^{2-}$ doping concentration $x$ could be as high as 75\%. All the ThCrAsN$_{1-x}$O$_x$ samples exhibit metallic electrical conduction. Nevertheless, none of the samples shows superconductivity above 30 mK, against the theoretical predictions.~\cite{LaCrAsO_2017_Pizarro,LaCrAsO_2017_Wang} Magnetic measurements and density functional theory (DFT) calculations indicate that ThCrAsN is G-type antiferromagnetic (AFM), the same as other reported Cr$_2$As$_2$-based compounds~\cite{LaCrAsO_2013,BaCrFeAs_2017,SrCr2As2_2017,EuCr2As2_2014}. And the magnetic structure of the hypothetical end member ThCrAsO, though unavailable, is also a G-type antiferromagnet within the DFT calculations, indicating the robustness of AFM order that hinder the appearance of superconductivity. From our results and analysis, the materials with Cr$_2$As$_2$ layers seem not to be a simple symmetry of FeSCs with respect to half 3$d$ shell filling.

	\section{Methods}
	\textit{Samples preparation.} Polycrystalline samples of ThCrAsN$_{1-x}$O$_x$ ($0\leq x\leq0.9$) were synthesized using powder of Th$_3$N$_4$, Th$_3$As$_4$, ThO$_2$, Cr and CrAs as starting materials. CrAs was prepared with Cr powder (99.99\%) and As pieces (99.999\%) and at 750$^\circ$C in evacuated quartz tubes. ThO$_2$ was heated to 700$^\circ$C for 12 h in the furnace to remove absorbed water. Preparation of the thorium metal ingot, Th$_3$N$_4$ powder and Th$_3$As$_4$ powder were described elsewhere~\cite{ThFeAsN,ThNiAsN,ThMnPN}. Stoichiometric mixture of the starting materials was ground and cold-pressed into a pellet. The pellet was loaded in an alumina crucible, which was sealed in an evacuated quartz tube. Subsequently, the tube was heated to 1100$^\circ$C in a muffle furnace, holding for 50 hours. The final product is dark grey and stable in air.
	
	\textit{Powder X-ray Diffraction.} Powder X-ray diffraction (XRD) experiments were carried out at room temperature on a PANalytical X-ray diffractometer with Cu $K_{\alpha1}$ radiation. XRD data were collected in the range $20^\circ\leq2\theta\leq150^\circ$ with a step of 0.013$^\circ$. The FullProf suite was used for the structural refinements~\cite{fullprof}.
	
	\textit{Resistivity and Magnetization Measurements.} A standard four-probe method was employed to collect the resistivity data. The data above 2 K were measured using a Quantum Design Physical Property Measurement System (PPMS) Dynacool, while the measurements between 30 mK and 0.8 K were performed in a dilution refrigerator. The direct-current (dc) magnetization was measured on a Quantum Design Magnetic Property Measurement System (MPMS3) equipped with the oven option.
	
	\textit{First-Principles Calculations.} The first-principles calculations were done within the generalized gradient approximation (GGA) by using the Vienna Ab-initio Simulation Package (VASP)~\cite{VASP}. The experimental crystal structure was used for the calculations. The plane-wave basis energy cutoff was chosen to be 550 eV. A $15\times15\times7$ $\Gamma$-centered K-mesh was used for the density-of-states (DOS) calculations. The Coulomb- and exchange parameters, $U$ and $J$, were introduced by using the GGA+$U$ calculations, where the parameters $U$ and $J$ are not independent and the difference ($U_\mathrm{eff}=U-J$) is meaningful. We adopted a GGA+$U$ ($U_\mathrm{eff}$ = 11 eV) correction on Th-$f$ shell to prevent unphysical 5$f$ component at Fermi level~\cite{LaCrAsO_2013,La_4f}. And $U_\mathrm{eff}$ was chosen at 0 eV for Cr 3$d$.

	\section{RESULTS AND DISCUSSION}
	
	\subsection{Crystal structure}
	
	\begin{figure*}
		\includegraphics[width=0.9\textwidth]{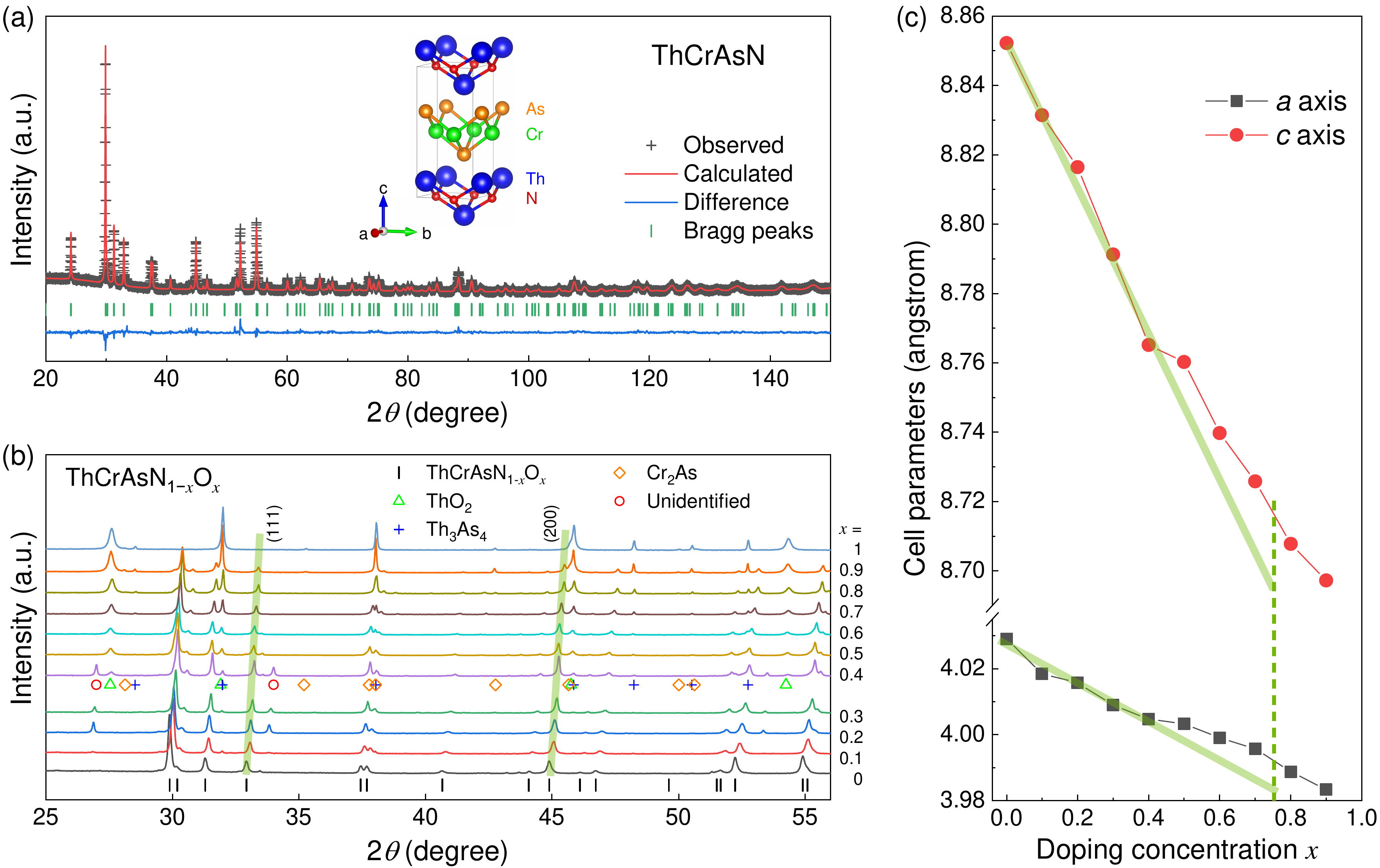}
		\caption{\label{Fig1_XRD}
			(a) Powder X-ray diffraction pattern and the corresponding Rietveld refinement for ThCrAsN, whose crystal structure is shown in the inset. (b) A series of XRD patterns from ThCrAsN$_{1-x}$O$_x$ ($x=0\sim1$) polycrystalline samples. The doping concentration $x$ for each pattern increases from bottom to top. The triangles, crosses, diamonds, and circles below the curve of $x=0.4$ mark the peaks from ThO$_2$, Th$_3$As$_4$, Cr$_2$As, and some unidentified impurities, respectively. The $hkl$ indices in space group $P4/nmm$ are shown with black tics below the curve of $x=0$. Two green bars are sketched to show the shift of characteristic peaks of ThCrAsN$_{1-x}$O$_x$. Note that there is no phase in space group $P4/nmm$ for the fully doped sample (the top curve, $x=1$). (c) Cell parameters of ThCrAsN$_{1-x}$O$_x$ as a function of the nominal doping concentration $x$ ($0\leq x\leq0.9$). The errors of data points are of the order of $10^{-4}$ \AA, which is invisible in comparison with the lattice parameters. So the error bars are not present. Two green bars are sketched based on the data points with $x\leq0.4$, and the green dash line is at $x=0.75$.
		}
	\end{figure*}

	\begin{table}
		\caption{\label{Tab1} Crystallographic data of ThCrAsN at 300 K obtained by the Rietveld refinement shown in Fig. \ref{Fig1_XRD}(a). The space group is $P4/nmm$ (No. 129). The occupancy of each atom was fixed to be 1.0, and the temperature factors were fixed to avoid unphysical negative values. Selected structural parameters of LaCrAsO are also listed for comparison~\cite{LaCrAsO_2013}. $h_{\mathrm{As}}$ in the table denotes the height of As from Cr plane.}
		\begin{ruledtabular}
			\begin{tabular}{cccccc}	
				Atom & Wyckoff & $x$ & $y$ & $z$ & $B_\mathrm{iso} (\mathrm{\AA}^{-2})$\\
				\hline
				Th & $2c$ & 0.25 & 0.25 & 0.1341(1) & 0.1\\
				Cr & $2a$ & 0.75  & 0.25 & 0.5 & 0.3\\
				As & $2c$ & 0.25 & 0.25 & 0.6673(3) & 0.3\\
				N & $2b$ & 0.75 & 0.25 & 0 & 1\\			
				\hline		
				\multicolumn{2}{l}{Compounds} & \multicolumn{2}{r}{ThCrAsN} & \multicolumn{2}{r}{LaCrAsO}~\cite{LaCrAsO_2013}  \\
				\hline
				\multicolumn{6}{l}{\textbf{Lattice parameters}}\\
				\multicolumn{2}{l}{$a$ (\AA)} &  \multicolumn{2}{r}{4.0290(1)} & \multicolumn{2}{r}{4.0412(3)}\\ 
				\multicolumn{2}{l}{$c$ (\AA)} & \multicolumn{2}{r}{8.8522(3)}  & \multicolumn{2}{r}{8.9863(7)} \\
				\multicolumn{2}{l}{$V$ (\AA$^3$)}&\multicolumn{2}{r}{143.70(1)}  & \multicolumn{2}{r}{146.76(2)} \\
				\multicolumn{2}{l}{$c/a$} &\multicolumn{2}{r}{2.197} & \multicolumn{2}{r}{2.224} \\
				\multicolumn{6}{l}{\textbf{Selected distances}}\\
				\multicolumn{2}{l}{$h_{\mathrm{As}}$ (\AA)} & \multicolumn{2}{r}{1.481(3)} & \multicolumn{2}{r}{1.460(2)}  \\
				\multicolumn{2}{l}{$d_\mathrm{Th-N/La-O}$ (\AA)} & \multicolumn{2}{r}{2.338(1)} & \multicolumn{2}{r}{2.364(1)} \\
				\multicolumn{2}{l}{$d_{\mathrm{Cr}-\mathrm{As}}$ (\AA)} &  \multicolumn{2}{r}{2.500(2)} & \multicolumn{2}{r}{2.494(1)} \\
				\multicolumn{2}{l}{$d_{\mathrm{Cr}-\mathrm{Cr}}$ (\AA)} &  \multicolumn{2}{r}{2.8489(1)} & \multicolumn{2}{r}{2.8576(3)} \\
				\multicolumn{6}{l}{\textbf{Bond angle}}\\
				\multicolumn{2}{l}{As$-$Cr$-$As ($^\circ$)} &  \multicolumn{2}{r}{107.36(10)}  &  \multicolumn{2}{r}{108.26(8)}  \\
			\end{tabular}
		\end{ruledtabular}
	\end{table}
	
	The XRD patterns of ThCrAsN and its oxygen-doped variants are displayed in Figs. \ref{Fig1_XRD}(a) and (b). The pattern of ThCrAsN can be indexed well using space group $P4/nmm$, indicating the successful preparation of the parent compound. Element substitutions of oxygen for nitrogen are carried out, ranging from 10\% oxygen to 100\% oxygen, and the doping step is equal to 10\%. With increasing oxygen concentration, the impurity peaks emerge and become conspicuous. The impurities of Th$_3$As$_4$ (for $x\geq0.1$), ThO$_2$ (for $x\geq0.4$), and Cr$_2$As (for $x\geq0.7$) can be identified, as well as some unknown impurity phase(s) (only for $0.2\leq x\leq 0.4$). However, ThCrAsN$_{1-x}$O$_x$ remains the main phase even though the nominal concentration of oxygen is as high as 80\%. Even for the pattern of $x=0.9$, the phase of 1111 is conspicuous enough to examine the existence of superconductivity. A simple estimation of the ratio between the 1111 phase and ThO$_2$ is presented in Supplemental Materials (SM)~\cite{suppmatt}. With regard to the wholly doped sample, i.e. ThCrAsO, no ZrCuSiAs-type phase can be identified. It's proper to believe ThCrAsO can not be synthesized under the present synthesis conditions.
	
	In Fig. \ref{Fig1_XRD}(b), two tilted green bars are sketched to show the monotonic shift of the (111) peaks and (200) peaks, suggesting the unit cell of ThCrAsN$_{1-x}$O$_x$ decreases gradually with the increase of oxygen doping. By a least-squares fit for the XRD patterns, the lattice parameters of ThCrAsN$_{1-x}$O$_x$ are determined and plotted as functions of nominal oxygen concentration $x$ in Fig. \ref{Fig1_XRD}(c). Both the $a$-axis and $c$-axis decrease almost linearly with increasing $x$ when $x\leq0.4$, as indicated by the green bars on the data points. The dependences of cell parameters on $x$ deviate from the green bars gradually when $x\geq0.5$, indicating the real oxygen concentration is lower than the nominal doping $x$, in line with the increasing impurity of ThO$_2$ in panel (b). If we take the slopes of $a(x)$ and $c(x)$ ($x\leq0.4$) as reference, the real oxygen concentration for the sample of ThCrAsN$_{0.1}$O$_{0.9}$ is about 0.75, inferred from its cell parameters. It is worth noting that the change of the $c$-axis is about 1.75\% from $x=0$ to 0.9, while the $a$-axis only shrinks by 1.13\%. That the $c$-axis changes much faster than the $a$-axis is consistent with the enhanced inter-layer coupling by electron doping.

	A Rietveld refinement was carried out using the collected XRD data shown in Fig. \ref{Fig1_XRD}(a). The refinement yields a weighted reliable factor of $R_\mathrm{wp} = 7.41\%$ and a goodness-of-fit of $S=1.54$, indicating reliability of the refinement. The resulting crystallographic data were summarized in Table \ref{Tab1}, and compared with the selected structural parameters of LaCrAsO. The axial ratio $c/a$ of ThCrAsN is smaller than that of LaCrAsO, implying ThCrAsN bears stronger internal chemical pressure along the $c$-axis. Similar reduction of $c/a$ was also observed in other siblings Th$M$AsN and La$M$AsO ($M$ = Mn, Fe, Ni, Co)~\cite{ThFeAsN,ThMnPN,ThNiAsN,ThCoAsN}. Other structural parameters of ThCrAsN, such as the height of As from Cr plane and As$-$Cr$-$As bond angle, are not distinct from those of LaCrAsO, which accounts for their close properties.

	\subsection{Electrical resistivity}
	\begin{figure}
		\includegraphics[width=0.45\textwidth]{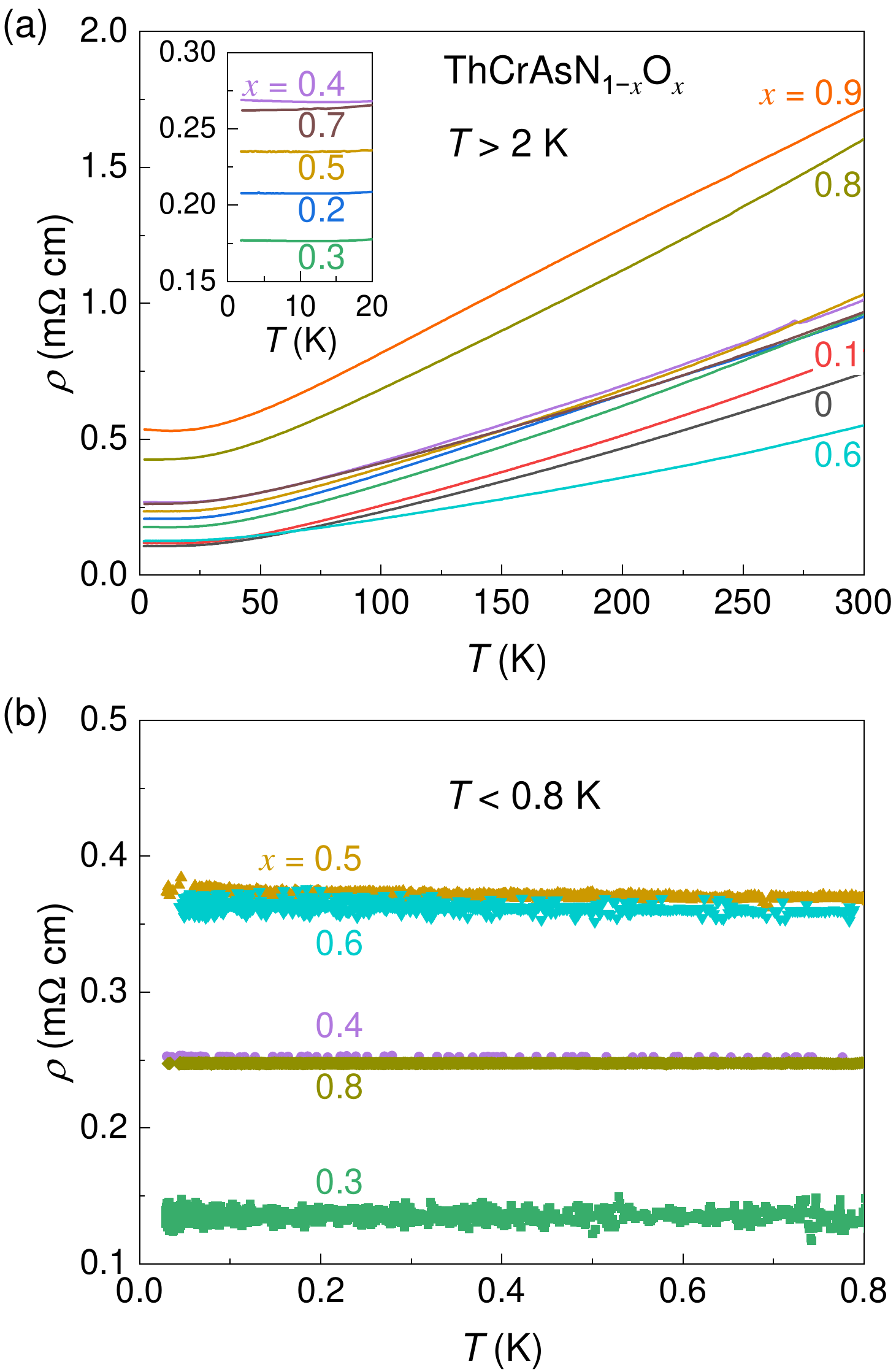}
		\caption{\label{Fig2_RT}
			Temperature dependences of resistivity ($\rho$) for ThCrAsN$_{1-x}$O$_x$. (a) $\rho(T)$ curves with $T>2$ K. The inset zooms in the data between 2 and 20 K for $x=$ 0.2, 0.3, 0.4, 0.5, and 0.7. (b) Data of $\rho(T)$ with 30 mK $< T <$ 0.8 K collected in the dilution refrigerator. The specimens with the same composition exhibit different residual resistivity in panels (a) and (b), due to variations in contact resistance and electrode size across different measurements.
		}
	\end{figure}
	
	The resistivity data ($\rho (T)$) for ThCrAsN$_{1-x}$O$_x$ are plotted in Fig. \ref{Fig2_RT}.  As seen in Fig. \ref{Fig2_RT}(a), $\rho (T)$ curves show a similar metallic behavior, regardless of the doping concentration. The magnitudes of resistivity are about 1 m$\Omega$ cm at room temperature, comparable to that of LaCrAsO~\cite{LaCrAsO_2013}. Only the samples of $x$ = 0.8 and 0.9 show a slightly higher resistivity due to the increased impurity proportion. Although the measured values of resistivity are affected by the contacts and impurity, the residual resistivity ratios (RRR) for all samples are between 2 and 3. The data below 20 K for $x=$ 0.2, 0.3, 0.4, 0.5, and 0.7 are enlarged in the inset of Fig. \ref{Fig2_RT}(a) so that the curves differentiate from each other. It's easily noticed that all curves but $x=0.9$ approach to constant values at 2 K. To examine the possibility that superconductivity emerges under lower temperatures, we measured $\rho (T)$ (30 mK $< T <$ 0.8 K) of several selected compounds in the dilution refrigerator, which are displayed in Fig. \ref{Fig2_RT}(b). However, the data of $\rho (T)$ are almost constant below 0.8 K and show no signs for superconductivity. Hence, superconductivity can't be induced in ThCrAsN$_{1-x}$O$_x$ even thogh the real electron doping concentration is as high as 0.75 ($n=4.75$), against the prediction of a superconductivity phase at $n>4.2$~\cite{LaCrAsO_2017_Wang}.

	\subsection{Magnetic properties}
	\begin{figure}
		\includegraphics[width=0.45\textwidth]{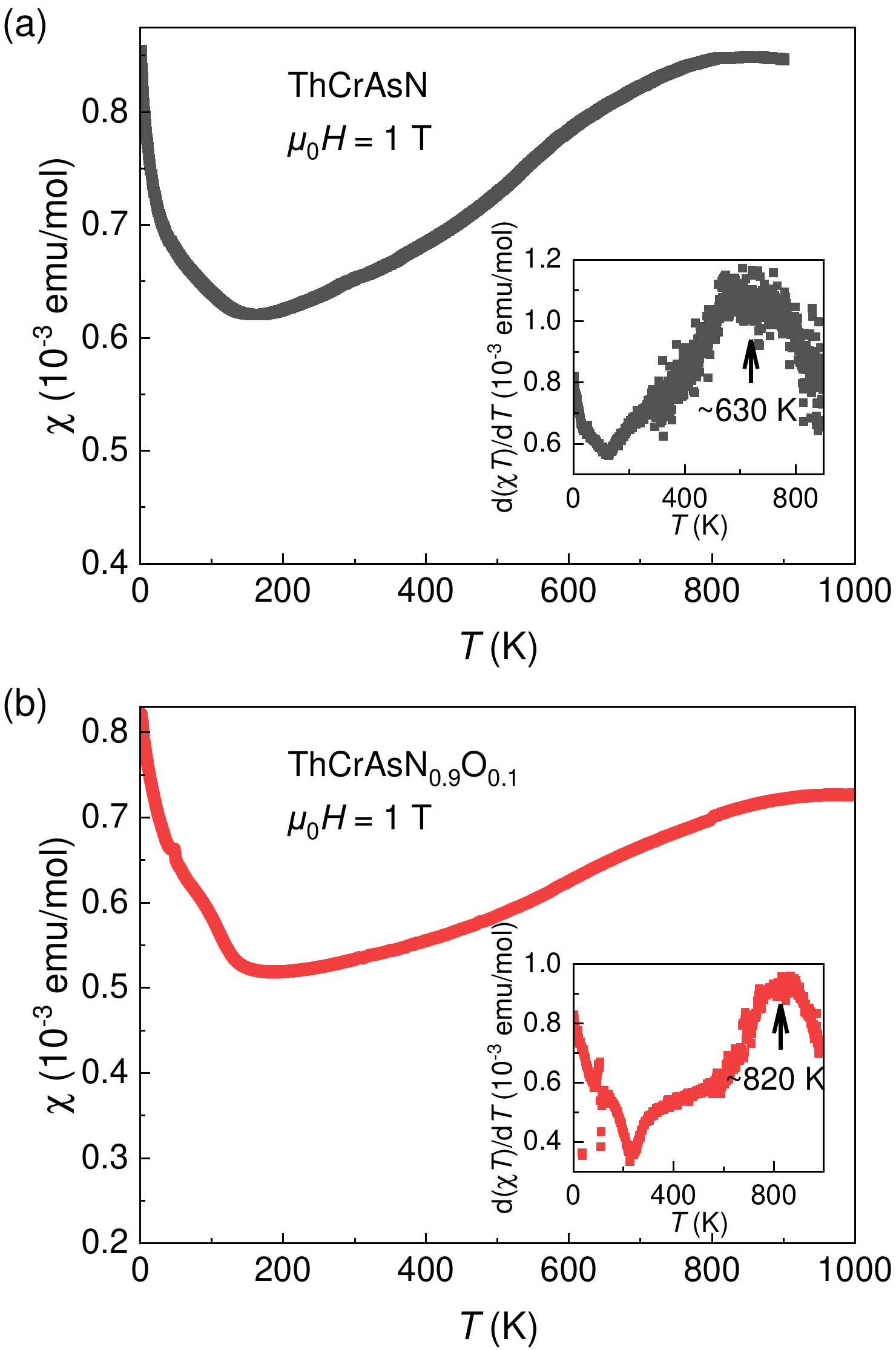}
		\caption{\label{Fig3_Magnetism}
			Temperature dependences of the magnetic susceptibility ($\chi$) measured at 1 T for (a) ThCrAsN and (b) ThCrAsN$ _{0.9} $O$ _{0.1} $. The small kink around 50 K in panel (b) is due to the oxygen contamination. $\mathrm{d}(\chi T)/\mathrm{d}T$ is plotted as a function of $T$ in the insets. 
		}
	\end{figure}
	Figure \ref{Fig3_Magnetism}(a) shows the $\chi(T)$ curve for the parent compound ThCrAsN under the magnetic field 1 T and there is no difference between zero-field-cooling and field-cooling data. Below 160 K, the susceptibility shows a Curie-Weiss-like tail, which may be caused by trace amount of paramagnetic impurity. Above 160 K, the susceptibility increases with the temperature monotonically and shows a gentle slope around 850 K, suggesting ThCrAsN is an antiferromagnet. The $\chi(T)$ behavior of ThCrAsN is reminiscent of the cases for polycrystalline SrCr$_2$As$_2$, which also shows no Curie-Weiss behavior below 900 K and no clear AFM transition~\cite{SrCr2As2_2017}. In addition, LaCrAsO only shows a smooth maximum at $520\sim570$ K~\cite{LaCrAsO_2013}. The behaviors of $\chi(T)$ for these Cr$_2$As$_2$-based materials are considered as the character of a two-dimensional (2D) antiferromagnet, which means that strong 2D AFM correlations may set in well above the ordering temperature ($T_N$)~\cite{LaCrAsO_2013}. Since the actual $T_N$ could be notably lower than the maximum in the susceptibility for 2D AFM compounds, here we use the derivative $\mathrm{d}(\chi T)/\mathrm{d}T$ to extract $T_N$ of ThCrAsN, as plotted in the inset~\cite{T_N,T_N_2,SrCr2As2_2017}. $\mathrm{d}(\chi T)/\mathrm{d}T$ shows an AFM transition peak at about 630 K, indicating $T_N$ of ThCrAsN is fall on same level with most Cr$_2$As$_2$-based materials~\cite{SrCr2As2_2017,BaCrFeAs_2017,EuCr2As2_2016,LaCrAsO_2013,Sr2Cr3As3O2_2018}.
	
	$\chi(T)$ for ThCrAsN$ _{0.9} $O$ _{0.1} $ under 1 T is displayed in Fig. \ref{Fig3_Magnetism}(b), which is quite akin to that of ThCrAsN. Yet the AFM order wasn't suppressed by the oxygen doping. As shown in the inset of Fig. \ref{Fig3_Magnetism}(b), $T_N$ of ThCrAsN$ _{0.9} $O$ _{0.1} $ is determined to be $\sim820$ K from $\mathrm{d}(\chi T)/\mathrm{d}T$, even higher than the parent. In addition, the susceptibility maximum of ThCrAsN$_{0.9}$O$_{0.1}$ appears at higher temperature than ThCrAsN, indicative of stronger 2D correlations. $\chi(T)$ curves for other ThCrAsN$_{1-x}$O$_x$ samples are shown in Fig. S2 in SM~\cite{suppmatt}. Unfortunately, the susceptibility data for other samples are not informative because of the strong background from the robust magnetic impurities. Nevertheless, we can still draw the conclusion that the susceptibility data show no sign of Meissner effect in ThCrAsN$_{1-x}$O$_x$.


	\subsection{Magnetic energies}
	\begin{figure}
		\includegraphics[width=0.45\textwidth]{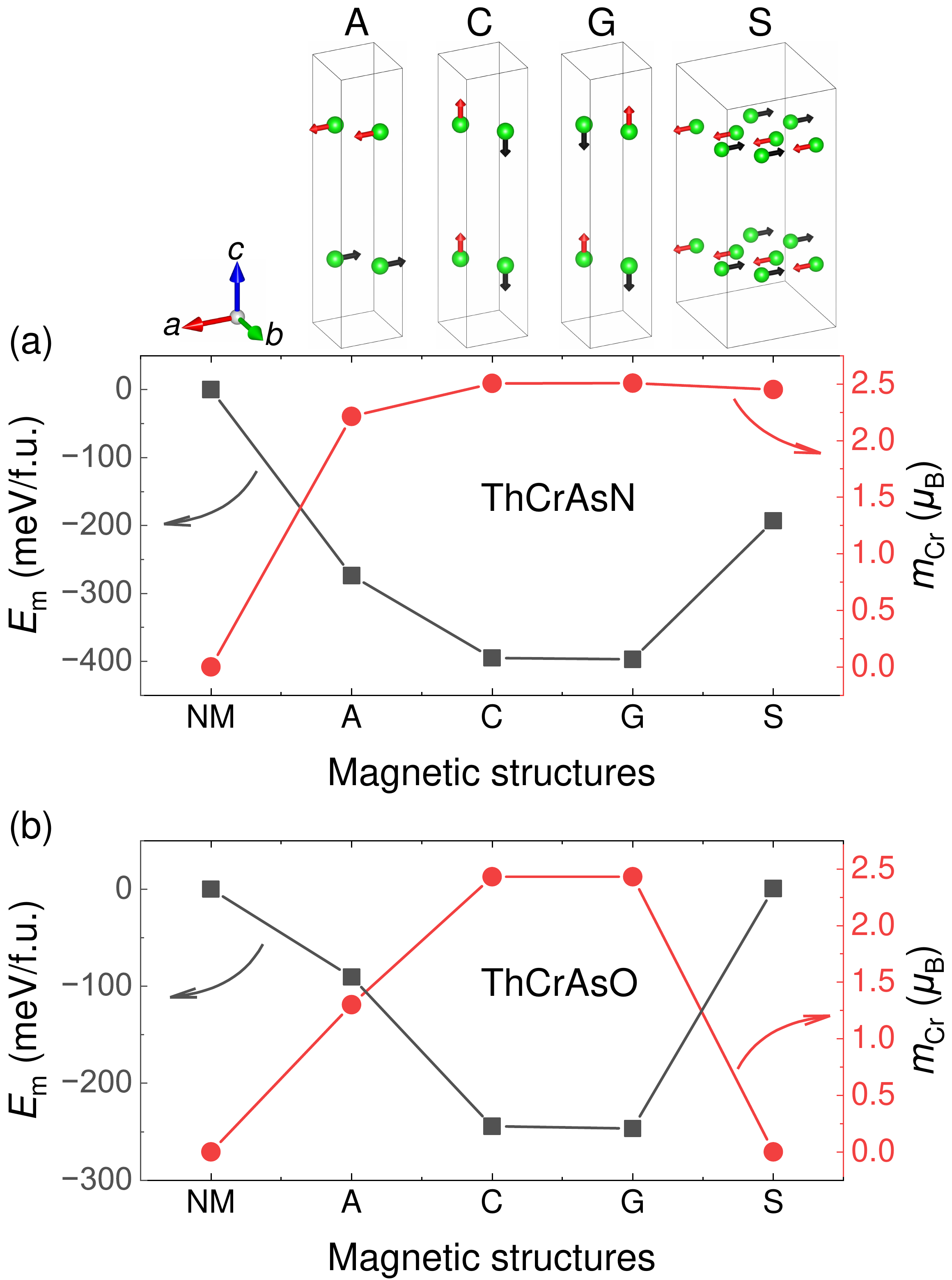}
		\caption{\label{Fig4_MagneticStructure}
			Calculated magnetic energies (black squares, left axis) and Cr spin moments (red circles, right axis) for (a) ThCrAsN and (b) ThCrAsO with different spin configurations. The notations for the magnetic structures are as follows: NM denotes non-magnetic states; A denotes A-type AFM order (in-plane FM order and out-of-plane AFM coupling); C denotes C-type AFM order (in-plane checkerboard AFM order and out-of-plane FM coupling); G denotes G-type AFM order (in-plane checkerboard AFM order and out-of-plane AFM coupling); S denotes the structure with in-plane striped AFM order and out-of-plane FM coupling. The spin configurations are presented on the top of the figure.
		}
	\end{figure}
	
	To determine the magnetic structure of ThCrAsN and its oxygen-doped derivatives, we examined the magnetic energies ($E_m$) for several possible AFM structures by performing DFT calculations. $E_m$ is defined as the energy difference between the spin-polarized state and nonmagnetic state. The selected spin configurations are presented in Fig. \ref{Fig4_MagneticStructure}, as well as the corresponding $E_m$ and Cr spin moments. $E_m$ and Cr spin moments are calculated for both ThCrAsN and ThCrAsO, in spite of the failure of synthesis for the latter.

	As shown in Fig. \ref{Fig4_MagneticStructure}(a), all magnetic orders lower the total energy of ThCrAsN, compared to the nonmagnetic state. Among them G-type AFM order is the most stable magnetic structure, yet C-type spin configuration has a pretty close energy. The similar $E_m$ values for G-type and C-type AFM orders demonstrate the interplane spin coupling for ThCrAsN is weak, consistent with the characteristic of 2D AFM order. Note that the calculated $E_m$ of G-type and C-type AFM orders for LaCrAsO are also reported to be close~\cite{LaCrAsO_2013}. And the spin structure of LaCrAsO is experimentally determined to be G-type. Based on the comparable behaviors of $\chi-T$ and results of $E_m$ calculations, it's plausible to anticipate that ThCrAsN has a G-type order, like LaCrAsO. In addition, the G-type AFM order were found to be the ground state for Cr-based 122-type coumpouds like BaCr$_2$As$_2$, BaCr$_2$P$_2$, SrCr$_2$As$_2$, EuCr$_2$As$_2$, Sr$_2$Cr$_3$As$_2$O$_2$, and Sr$_2$Cr$_2$AsO$_3$~\cite{BaCrFeAs_2017,SrCr2As2_2017,EuCr2As2_2014,BaCr2P2_2019,Sr2Cr3As3O2,Sr2Cr2AsO3}, which also supports our speculation of the magnetic structure of ThCrAsN. 
	
	To gain a deeper insight into the magnetic ordering of ThCrAsN, we calculate the exchange interactions between the Cr neighbors in light of a simple Heisenberg model~\cite{Heisenberg_model}. The magnetic energies of C-type, G-type and S-type AFM orderings can be expressed as
	\begin{eqnarray}
		E_\mathrm{C}=(-2J_1+2J_2+J_c)S^2,\\
		E_\mathrm{G}=(-2J_1+2J_2-J_c)S^2,\\
		E_\mathrm{S}=(-2J_2+J_c)S^2,
	\end{eqnarray}
	where $S$ is the local spin, and $J_1/J_2/J_c$ refer to the interactions of in-plane nearest-neighbor, in-plane next-nearest-neighbor, out-of-plane nearest-neighbor, respectively. The values of energies for ThCrAsN could be found in the Table S2 in the SM~\cite{suppmatt}. Solutions of the equations are 
	\begin{eqnarray}
		J_1=(-E_\mathrm{G}-E_\mathrm{S})/2S^2,\\ J_2=(E_\mathrm{C}-E_\mathrm{G}-2E_\mathrm{S})/4S^2,\\ J_c=(E_\mathrm{C}-E_\mathrm{G})/2S^2.
	\end{eqnarray}
	 Assuming the Cr spin moment of ThCrAsN is 2.5 $\mu_B$ ($S=1.25$, according to Fig. \ref{Fig4_MagneticStructure}(a)),  the resulting interactions are $ J_1 $ = 187.9 meV, $ J_2 $ = 62.1 meV, $ J_c $ = 0.66 meV, consistent with the stability conditions of G-type AFM order that $ J_1>0$, $ J_1>2J_2 $, and $ J_c>0 $. The small positive $J_c$ value confirms the AFM coupling between the adjacent Cr$_2$As$_2$ layers is weak, in good agreement with the $\chi(T)$ behavior. The large positive values of $J_1$ and $J_1/J_2$ indicate that in-plane Cr--Cr direct interaction is dominant, thus the high Neel temperature and robust G-type AFM order is explicit. We also notice that the calculated energies $E_\mathrm{C}$, $E_\mathrm{G}$, and $E_\mathrm{S}$ for ThCrAsN, as well as the Cr moment, are all close to the values for LaCrAsO~\cite{LaCrAsO_2013}. Hence the resultant $J_1$ and $J_2$ are almost the same as the interactions for ThCrAsN.

	$E_m$ and Cr spin moments of ThCrAsO ($3d^5$ for Cr, one electron more compared to ThCrAsN) with the same magnetic structures as ThCrAsN are shown in Fig. \ref{Fig4_MagneticStructure}(b). The calculation indicates that G-type order is still the ground state of ThCrAsO, although its $E_m$ is significantly lower than ThCrAsN. And the corresponding Cr spin moment is reduced slightly to be 2.43 $\mu_\mathrm{B}$. That is to say, AFM order of ThCrAsN is to a certain extent suppressed by electron doping. However, the G-type magnetic order remains for the end member ThCrAsO.

	Here we would like to point out that $J_1$ and $J_2$ values for LaFeAsO both are about 50 meV/$S^2$, which are close and conspicuously smaller than the values for ThCrAsN~\cite{LaFeAsO_J}. The competition of exchange interactions in FeSCs lead to the collinear (or bicollinear) AFM order, contrary to the case for ThCrAsN. And the next-nearest-neighbor interaction $J_2$ is generally believed to play an important role in the electron pairing mechanism of FeSCs~\cite{J2_PRB,J2_SR}. The magnetic structures and exchange interactions clearly point to the distinction between Cr$_2$As$_2$-based materials and FeSCs.

	\subsection{Density of states}
	\begin{figure*}
		\includegraphics[width=\textwidth]{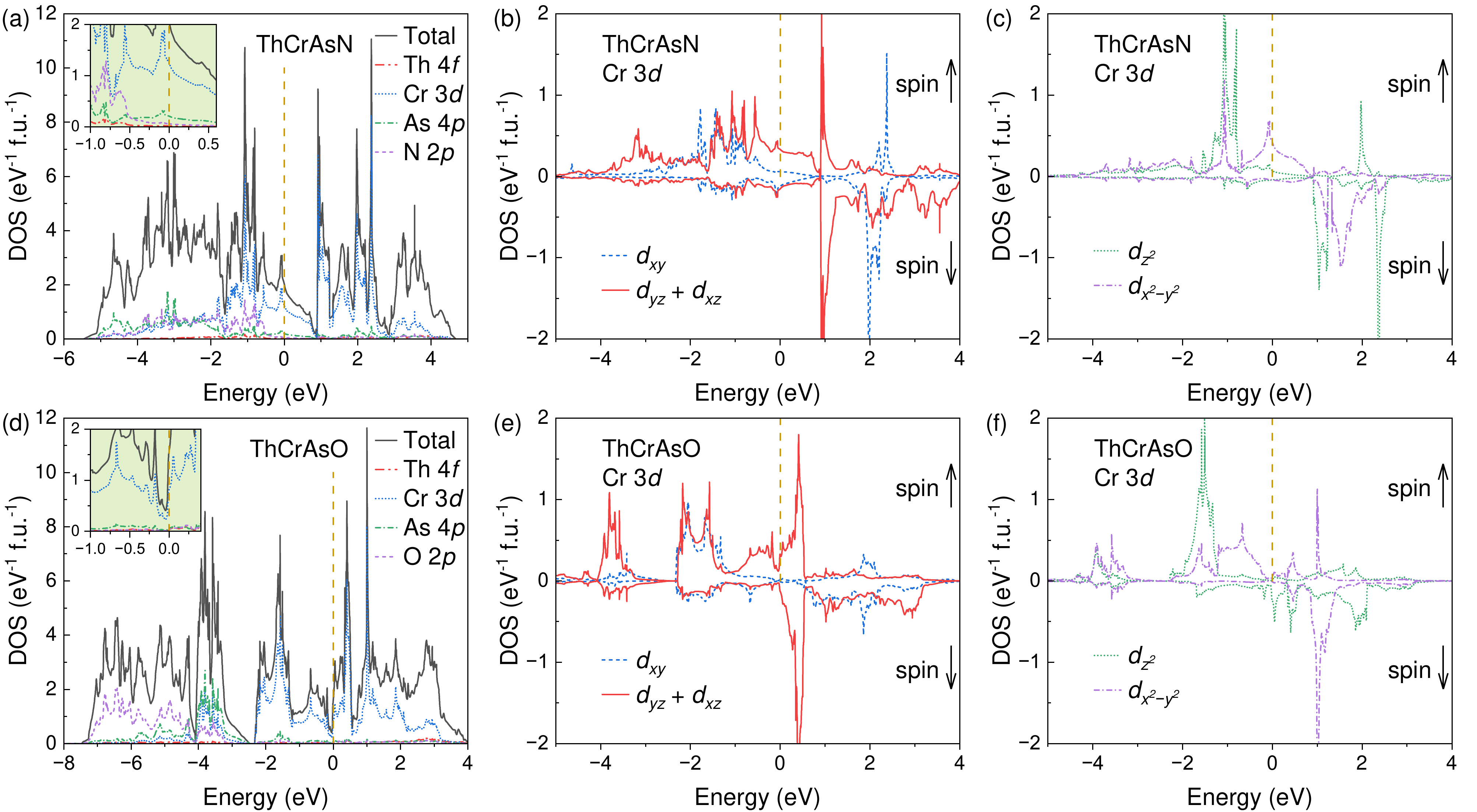}
		\caption{\label{Fig5_DOS}
			[(a), (d)] Total electronic density of states of G-type AFM ordered ThCrAsN and ThCrAsO, respectively. Insets of (a) and (d) zoom in on the DOS around the Fermi energy. [(b), (c), (e), (f)] Projected density of states of 3$d$ orbitals of Cr in the ThCrAsN and ThCrAsO, respectively. The contributions of $d_{xz}$ and $d_{yz}$ are combined because they are degenerate.
		}
	\end{figure*}
	
	The total density of states (DOS) of G-type AFM ordered ThCrAsN and ThCrAsO are calculated and plotted in Fig. \ref{Fig5_DOS}, as well as their projected density of states (PDOS) of Cr-3$d$ orbitals. Both ThCrAsN and ThCrAsO show a relatively large DOS at the Fermi energy ($E_\mathrm{F}$), coinciding with the typical metallic behavior of ThCrAsN$_{1-x}$O$_x$. As shown in Fig. \ref{Fig5_DOS}(a), the DOS of ThCrAsN around $E_\mathrm{F}$ is mainly contributed by Cr-3$d$ and As-4$p$ orbitals. And it is easy to recognize the modest hybridizations between Cr-3$d$ and As-4$p$ in the inset. The $d-p$ hybridizations account for the itinerant magnetism and the reduced Cr-3$d$ magnetic moment of ThCrAsN (theoretically 4 $\mu_\mathrm{B}$ for Cr$^{2+}$). 
	
	DOS of ThCrAsO is shown in Fig. \ref{Fig5_DOS}(d). DOSs for ThCrAsN and ThCrAsO are largely comparable, and $E_\mathrm{F}$ of ThCrAsO shifts to higher position due to its nominal Cr-3$d^5$ electron configuration. The similarities of DOSs can be seen better in other panels about PDOS of Cr. Cr-3$d$ electrons still dominate the states around $E_\mathrm{F}$ of ThCrAsO, but the hybridizations between Cr-3$d$ and As-4$p$ is negligible compared to ThCrAsN.

	To compare the PDOS of Cr-3$d$ orbitals for ThCrAsN and ThCrAsO more clearly, we divide the orbitals into two groups and present the curves in Figs. \ref{Fig5_DOS}(b)(e) ($d_{xy}$, $d_{yz}+d_{xz}$) and (c)(f) ($d_{z^2}$, $d_{x^2-y^2}$), respectively. The itinerant magnetism of ThFeAsN and ThFeAsO is confirmed by the broad energy bands. We notice that the orbital occupations of Cr for both ThCrAsN and ThCrAsO are highly spin-polarized, indicating the crystal field splitting energy is small so that Hund's rule is followed. The high-spin state is also observed for LaCrAsO and LaMnAsO, while the spin polarization for LaFeAsO is much weaker~\cite{LaCrAsO_2013,LaFeAsO_J}. For ThCrAsN, the PDOS at $E_\mathrm{F}$ comes from all five 3$d$ orbitals, and the $d_{x^2-y^2}$ and $d_{yz}+d_{xz}$ contribute substantially. The heavy proportion of $d_{x^2-y^2}$ orbital means enhanced direct exchange at $E_\mathrm{F}$, consistent with the G-type AFM order that is favored by the nearest Cr--Cr interactions. After the replacement of nitrogen by oxygen, the DOS of $d_{x^2-y^2}$ and $d_{xy}$ decline sharply, meanwhile the contributions from $d_{z^2}$ increase significantly, preceded only by $d_{yz}+d_{xz}$. Whether for ThCrAsN or ThCrAsO, there is a heavy mixing of the $t_{2}$ and $e$ orbitals at $E_\mathrm{F}$, which is deemed detrimental to high-$T_c$ superconductivity in some theoretical work~\cite{HuJP_2015}.
	
	\section{Concluding Remarks}
	 	
	To summarize, we have successfully synthesized a new 1111-type material, ThCrAsN. Then we explored the possibility of superconductivity by electron-doping in ThCrAsN through O$^{2-}$ substitution for N$^{3-}$. However, ThCrAsN$_{1-x}$O$_x$ shows no sign of superconductivity with the oxygen concentration from 0 to 75\%, inconsistent with the theoretical predictions. Magnetic susceptibility of ThCrAsN indicates its AFM ground states with an ordering temperature about 630 K, and DFT calculations imply the most probable spin structure for ThCrAsN and the doped variants is G-type, the same as LaCrAsO and $A$Cr$_2$As$_2$ ($A$ = Sr, Ba, Eu), although further examination of the magnetic structure with powder neutron scattering is still required. The analysis of the exchange interactions between Cr neighbors indicates in-plane direct interaction $J_1$ is dominant, resulting in the robust G-type AFM order of ThCrAs(N,O). Although the superconductivity couldn't be induced by electron doping in ThCrAsN, the AFM order is moderately suppressed, manifested by the reduced magnetic energy and Cr spin moment of ThCrAsO. In addition, the PDOSs of Cr show that the spin polarization for ThCrAsN and ThCrAsO is strong, which is a major difference from the DOS of LaFeAsO.
	
	Since the pairing mechanism in FeSCs is still under debate, it's hard to elucidate the reasons for absence of superconductivity in electron-doped ThCrAsN. Nevertheless, the distinctions between Cr$_2$As$_2$-based materials and FeSCs are explicit. In general, high-$T_c$ superconductivity develops after a magnetic ordering is suppressed, suggesting the interplay between superconductivity and magnetism is crucial. G-type AFM order is usually found to be the ground state for Cr$_2$As$_2$-based compounds, which is apparently distinguished from collinear or bicollinear AFM orders for FeSCs. Essentially, the exchange interactions between Cr (or Fe) neighbors shape the magnetic structures and may be related to the superconducting pairing. And the different spin polarizations for FeSCs and Cr-based materials are also not inconsiderable. In addition, the electronic correlations may play an important role as well. In fact, an angle-resolved photoemission spectroscopy study shows that BaCr$_2$As$_2$ is much less correlated than BaFe$_2$As$_2$~\cite{BaCr2As2_ARPES}. The absence of superconductivity in Cr$_2$As$_2$-based materials calls for an indepth theoretical explanation.
		
	Although the comparability of Cr-based compounds and Fe-based counterparts is highlighted in some theoretical works based on the doped-Mott scenario, their similarities shouldn't be overestimated in respect of our work and other available experiments. Yet the possibility of superconductivity in Cr-based materials is not ruled out. We hope further experiments, like high-pressure measurements and hole doping in the synthesis, will suppress the magnetic ordering and give rise to superconductivity in Cr$_2$As$_2$-based materials.

	\begin{acknowledgments}
	This work was supported by the National Natural Science Foundation of China (Grants No. 12204094 and No. 12050003), the National Key Research and Development Program of China (Grant No. 2022YFA1403202), the Key Research and Development Program of Zhejiang Province, China (Grant No. 2021C01002), the Natural Science Foundation of Jiangsu Province (Grant No. BK20220796), and the open research fund of Key Laboratory of Quantum Materials and Devices (Southeast University), Ministry of Education.

	\end{acknowledgments}
	
	\bibliography{ThCrAsN1-xOx.bib}

\begin{thebibliography}{44}%
\makeatletter
\providecommand \@ifxundefined [1]{%
 \@ifx{#1\undefined}
}%
\providecommand \@ifnum [1]{%
 \ifnum #1\expandafter \@firstoftwo
 \else \expandafter \@secondoftwo
 \fi
}%
\providecommand \@ifx [1]{%
 \ifx #1\expandafter \@firstoftwo
 \else \expandafter \@secondoftwo
 \fi
}%
\providecommand \natexlab [1]{#1}%
\providecommand \enquote  [1]{``#1''}%
\providecommand \bibnamefont  [1]{#1}%
\providecommand \bibfnamefont [1]{#1}%
\providecommand \citenamefont [1]{#1}%
\providecommand \href@noop [0]{\@secondoftwo}%
\providecommand \href [0]{\begingroup \@sanitize@url \@href}%
\providecommand \@href[1]{\@@startlink{#1}\@@href}%
\providecommand \@@href[1]{\endgroup#1\@@endlink}%
\providecommand \@sanitize@url [0]{\catcode `\\12\catcode `\$12\catcode
  `\&12\catcode `\#12\catcode `\^12\catcode `\_12\catcode `\%12\relax}%
\providecommand \@@startlink[1]{}%
\providecommand \@@endlink[0]{}%
\providecommand \url  [0]{\begingroup\@sanitize@url \@url }%
\providecommand \@url [1]{\endgroup\@href {#1}{\urlprefix }}%
\providecommand \urlprefix  [0]{URL }%
\providecommand \Eprint [0]{\href }%
\providecommand \doibase [0]{https://doi.org/}%
\providecommand \selectlanguage [0]{\@gobble}%
\providecommand \bibinfo  [0]{\@secondoftwo}%
\providecommand \bibfield  [0]{\@secondoftwo}%
\providecommand \translation [1]{[#1]}%
\providecommand \BibitemOpen [0]{}%
\providecommand \bibitemStop [0]{}%
\providecommand \bibitemNoStop [0]{.\EOS\space}%
\providecommand \EOS [0]{\spacefactor3000\relax}%
\providecommand \BibitemShut  [1]{\csname bibitem#1\endcsname}%
\let\auto@bib@innerbib\@empty
\bibitem [{\citenamefont {Kamihara}\ \emph {et~al.}(2008)\citenamefont
  {Kamihara}, \citenamefont {Watanabe}, \citenamefont {Hirano},\ and\
  \citenamefont {Hosono}}]{LaFeAsO_2008}%
  \BibitemOpen
  \bibfield  {author} {\bibinfo {author} {\bibfnamefont {Y.}~\bibnamefont
  {Kamihara}}, \bibinfo {author} {\bibfnamefont {T.}~\bibnamefont {Watanabe}},
  \bibinfo {author} {\bibfnamefont {M.}~\bibnamefont {Hirano}},\ and\ \bibinfo
  {author} {\bibfnamefont {H.}~\bibnamefont {Hosono}},\ }\bibfield  {title}
  {\bibinfo {title} {Iron-based layered superconductor
  {La}[{O}$_{1-x}${F}$_x$]{FeAs} ($x= 0.05- 0.12$) with ${T}_c= 26$ {K}},\
  }\href {https://doi.org/10.1021/ja800073m} {\bibfield  {journal} {\bibinfo
  {journal} {J. Am. Chem. Soc.}\ }\textbf {\bibinfo {volume} {130}},\ \bibinfo
  {pages} {3296} (\bibinfo {year} {2008})}\BibitemShut {NoStop}%
\bibitem [{\citenamefont {Johrendt}\ \emph {et~al.}(2011)\citenamefont
  {Johrendt}, \citenamefont {Hosono}, \citenamefont {Hoffmann},\ and\
  \citenamefont {Pöttgen}}]{Review_2011}%
  \BibitemOpen
  \bibfield  {author} {\bibinfo {author} {\bibfnamefont {D.}~\bibnamefont
  {Johrendt}}, \bibinfo {author} {\bibfnamefont {H.}~\bibnamefont {Hosono}},
  \bibinfo {author} {\bibfnamefont {R.-D.}\ \bibnamefont {Hoffmann}},\ and\
  \bibinfo {author} {\bibfnamefont {R.}~\bibnamefont {Pöttgen}},\ }\bibfield
  {title} {\bibinfo {title} {Structural chemistry of superconducting pnictides
  and pnictide oxides with layered structures},\ }\href
  {https://doi.org/doi:10.1524/zkri.2011.1363} {\bibfield  {journal} {\bibinfo
  {journal} {Z. Kristallogr.}\ }\textbf {\bibinfo {volume} {226}},\ \bibinfo
  {pages} {435} (\bibinfo {year} {2011})}\BibitemShut {NoStop}%
\bibitem [{\citenamefont {Shatruk}(2019)}]{Review_2019}%
  \BibitemOpen
  \bibfield  {author} {\bibinfo {author} {\bibfnamefont {M.}~\bibnamefont
  {Shatruk}},\ }\bibfield  {title} {\bibinfo {title} {{ThCr}$_2${Si}$_2$
  structure type: The “perovskite” of intermetallics},\ }\href
  {https://doi.org/https://doi.org/10.1016/j.jssc.2019.02.012} {\bibfield
  {journal} {\bibinfo  {journal} {J. Solid State Chem.}\ }\textbf {\bibinfo
  {volume} {272}},\ \bibinfo {pages} {198} (\bibinfo {year}
  {2019})}\BibitemShut {NoStop}%
\bibitem [{\citenamefont {Park}\ \emph {et~al.}(2013)\citenamefont {Park},
  \citenamefont {Mizoguchi}, \citenamefont {Kodama}, \citenamefont {Shamoto},
  \citenamefont {Otomo}, \citenamefont {Matsuishi}, \citenamefont {Kamiya},\
  and\ \citenamefont {Hosono}}]{LaCrAsO_2013}%
  \BibitemOpen
  \bibfield  {author} {\bibinfo {author} {\bibfnamefont {S.-W.}\ \bibnamefont
  {Park}}, \bibinfo {author} {\bibfnamefont {H.}~\bibnamefont {Mizoguchi}},
  \bibinfo {author} {\bibfnamefont {K.}~\bibnamefont {Kodama}}, \bibinfo
  {author} {\bibfnamefont {S.-i.}\ \bibnamefont {Shamoto}}, \bibinfo {author}
  {\bibfnamefont {T.}~\bibnamefont {Otomo}}, \bibinfo {author} {\bibfnamefont
  {S.}~\bibnamefont {Matsuishi}}, \bibinfo {author} {\bibfnamefont
  {T.}~\bibnamefont {Kamiya}},\ and\ \bibinfo {author} {\bibfnamefont
  {H.}~\bibnamefont {Hosono}},\ }\bibfield  {title} {\bibinfo {title} {Magnetic
  {Structure} and {Electromagnetic} {Properties} of {LnCrAsO} with a
  {ZrCuSiAs}-type {Structure} ({Ln = La, Ce, Pr, and Nd})},\ }\href
  {https://doi.org/10.1021/ic401487q} {\bibfield  {journal} {\bibinfo
  {journal} {Inorg. Chem.}\ }\textbf {\bibinfo {volume} {52}},\ \bibinfo
  {pages} {13363} (\bibinfo {year} {2013})}\BibitemShut {NoStop}%
\bibitem [{\citenamefont {Singh}\ \emph {et~al.}(2009)\citenamefont {Singh},
  \citenamefont {Sefat}, \citenamefont {McGuire}, \citenamefont {Sales},
  \citenamefont {Mandrus}, \citenamefont {VanBebber},\ and\ \citenamefont
  {Keppens}}]{BaCr2As2_2009}%
  \BibitemOpen
  \bibfield  {author} {\bibinfo {author} {\bibfnamefont {D.~J.}\ \bibnamefont
  {Singh}}, \bibinfo {author} {\bibfnamefont {A.~S.}\ \bibnamefont {Sefat}},
  \bibinfo {author} {\bibfnamefont {M.~A.}\ \bibnamefont {McGuire}}, \bibinfo
  {author} {\bibfnamefont {B.~C.}\ \bibnamefont {Sales}}, \bibinfo {author}
  {\bibfnamefont {D.}~\bibnamefont {Mandrus}}, \bibinfo {author} {\bibfnamefont
  {L.~H.}\ \bibnamefont {VanBebber}},\ and\ \bibinfo {author} {\bibfnamefont
  {V.}~\bibnamefont {Keppens}},\ }\bibfield  {title} {\bibinfo {title}
  {Itinerant antiferromagnetism in {${\text{BaCr}}_{2}{\text{As}}_{2}$}:
  Experimental characterization and electronic structure calculations},\ }\href
  {https://doi.org/10.1103/PhysRevB.79.094429} {\bibfield  {journal} {\bibinfo
  {journal} {Phys. Rev. B}\ }\textbf {\bibinfo {volume} {79}},\ \bibinfo
  {pages} {094429} (\bibinfo {year} {2009})}\BibitemShut {NoStop}%
\bibitem [{\citenamefont {Paramanik}\ \emph {et~al.}(2014)\citenamefont
  {Paramanik}, \citenamefont {Prasad}, \citenamefont {Geibel},\ and\
  \citenamefont {Hossain}}]{EuCr2As2_2014}%
  \BibitemOpen
  \bibfield  {author} {\bibinfo {author} {\bibfnamefont {U.~B.}\ \bibnamefont
  {Paramanik}}, \bibinfo {author} {\bibfnamefont {R.}~\bibnamefont {Prasad}},
  \bibinfo {author} {\bibfnamefont {C.}~\bibnamefont {Geibel}},\ and\ \bibinfo
  {author} {\bibfnamefont {Z.}~\bibnamefont {Hossain}},\ }\bibfield  {title}
  {\bibinfo {title} {Itinerant and local-moment magnetism in
  {${\text{EuCr}}_{2}$${\text{As}}_{2}$} single crystals},\ }\href
  {https://doi.org/10.1103/PhysRevB.89.144423} {\bibfield  {journal} {\bibinfo
  {journal} {Phys. Rev. B}\ }\textbf {\bibinfo {volume} {89}},\ \bibinfo
  {pages} {144423} (\bibinfo {year} {2014})}\BibitemShut {NoStop}%
\bibitem [{\citenamefont {Das}\ \emph {et~al.}(2017)\citenamefont {Das},
  \citenamefont {Sangeetha}, \citenamefont {Lindemann}, \citenamefont
  {Heitmann}, \citenamefont {Kreyssig}, \citenamefont {Goldman}, \citenamefont
  {McQueeney}, \citenamefont {Johnston},\ and\ \citenamefont
  {Vaknin}}]{SrCr2As2_2017}%
  \BibitemOpen
  \bibfield  {author} {\bibinfo {author} {\bibfnamefont {P.}~\bibnamefont
  {Das}}, \bibinfo {author} {\bibfnamefont {N.~S.}\ \bibnamefont {Sangeetha}},
  \bibinfo {author} {\bibfnamefont {G.~R.}\ \bibnamefont {Lindemann}}, \bibinfo
  {author} {\bibfnamefont {T.~W.}\ \bibnamefont {Heitmann}}, \bibinfo {author}
  {\bibfnamefont {A.}~\bibnamefont {Kreyssig}}, \bibinfo {author}
  {\bibfnamefont {A.~I.}\ \bibnamefont {Goldman}}, \bibinfo {author}
  {\bibfnamefont {R.~J.}\ \bibnamefont {McQueeney}}, \bibinfo {author}
  {\bibfnamefont {D.~C.}\ \bibnamefont {Johnston}},\ and\ \bibinfo {author}
  {\bibfnamefont {D.}~\bibnamefont {Vaknin}},\ }\bibfield  {title} {\bibinfo
  {title} {Itinerant {G}-type antiferromagnetic order in
  {${\mathrm{SrCr}}_{2}{\mathrm{As}}_{2}$}},\ }\href
  {https://doi.org/10.1103/PhysRevB.96.014411} {\bibfield  {journal} {\bibinfo
  {journal} {Phys. Rev. B}\ }\textbf {\bibinfo {volume} {96}},\ \bibinfo
  {pages} {014411} (\bibinfo {year} {2017})}\BibitemShut {NoStop}%
\bibitem [{\citenamefont {Jiang}\ \emph {et~al.}(2015)\citenamefont {Jiang},
  \citenamefont {Bao}, \citenamefont {Zhai}, \citenamefont {Tang},
  \citenamefont {Sun}, \citenamefont {Liu}, \citenamefont {Wang}, \citenamefont
  {Bai}, \citenamefont {Xu},\ and\ \citenamefont {Cao}}]{Sr2Cr3As3O2}%
  \BibitemOpen
  \bibfield  {author} {\bibinfo {author} {\bibfnamefont {H.}~\bibnamefont
  {Jiang}}, \bibinfo {author} {\bibfnamefont {J.-K.}\ \bibnamefont {Bao}},
  \bibinfo {author} {\bibfnamefont {H.-F.}\ \bibnamefont {Zhai}}, \bibinfo
  {author} {\bibfnamefont {Z.-T.}\ \bibnamefont {Tang}}, \bibinfo {author}
  {\bibfnamefont {Y.-L.}\ \bibnamefont {Sun}}, \bibinfo {author} {\bibfnamefont
  {Y.}~\bibnamefont {Liu}}, \bibinfo {author} {\bibfnamefont {Z.-C.}\
  \bibnamefont {Wang}}, \bibinfo {author} {\bibfnamefont {H.}~\bibnamefont
  {Bai}}, \bibinfo {author} {\bibfnamefont {Z.-A.}\ \bibnamefont {Xu}},\ and\
  \bibinfo {author} {\bibfnamefont {G.-H.}\ \bibnamefont {Cao}},\ }\bibfield
  {title} {\bibinfo {title} {Physical properties and electronic structure of
  {${\mathrm{Sr}}_{2}{\mathrm{Cr}}_{3}{\mathrm{As}}_{2}{\mathrm{O}}_{2}$}
  containing {${\mathrm{CrO}}_{2}$} and {${\mathrm{Cr}}_{2}{\mathrm{As}}_{2}$}
  square-planar lattices},\ }\href {https://doi.org/10.1103/PhysRevB.92.205107}
  {\bibfield  {journal} {\bibinfo  {journal} {Phys. Rev. B}\ }\textbf {\bibinfo
  {volume} {92}},\ \bibinfo {pages} {205107} (\bibinfo {year}
  {2015})}\BibitemShut {NoStop}%
\bibitem [{\citenamefont {Pavan Kumar~Naik}\ \emph {et~al.}(2021)\citenamefont
  {Pavan Kumar~Naik}, \citenamefont {Iwasa}, \citenamefont {Kuramochi},
  \citenamefont {Ichihara}, \citenamefont {Kishio}, \citenamefont {Hongo},
  \citenamefont {Maezono}, \citenamefont {Nishio},\ and\ \citenamefont
  {Ogino}}]{Sr2ScCrAsO3}%
  \BibitemOpen
  \bibfield  {author} {\bibinfo {author} {\bibfnamefont {S.}~\bibnamefont
  {Pavan Kumar~Naik}}, \bibinfo {author} {\bibfnamefont {Y.}~\bibnamefont
  {Iwasa}}, \bibinfo {author} {\bibfnamefont {K.}~\bibnamefont {Kuramochi}},
  \bibinfo {author} {\bibfnamefont {Y.}~\bibnamefont {Ichihara}}, \bibinfo
  {author} {\bibfnamefont {K.}~\bibnamefont {Kishio}}, \bibinfo {author}
  {\bibfnamefont {K.}~\bibnamefont {Hongo}}, \bibinfo {author} {\bibfnamefont
  {R.}~\bibnamefont {Maezono}}, \bibinfo {author} {\bibfnamefont
  {T.}~\bibnamefont {Nishio}},\ and\ \bibinfo {author} {\bibfnamefont
  {H.}~\bibnamefont {Ogino}},\ }\bibfield  {title} {\bibinfo {title}
  {Synthesis, electronic structure, and physical properties of layered
  oxypnictides {Sr$_2$ScCrAsO$_3$} and {Ba$_3$Sc$_2$Cr$_2$As$_2$O$_5$}},\
  }\href {https://doi.org/10.1021/acs.inorgchem.0c03404} {\bibfield  {journal}
  {\bibinfo  {journal} {Inorg. Chem.}\ }\textbf {\bibinfo {volume} {60}},\
  \bibinfo {pages} {1930} (\bibinfo {year} {2021})}\BibitemShut {NoStop}%
\bibitem [{\citenamefont {Ablimit}\ \emph {et~al.}(2017)\citenamefont
  {Ablimit}, \citenamefont {Sun}, \citenamefont {Jiang}, \citenamefont {Bao},
  \citenamefont {Zhai}, \citenamefont {Tang}, \citenamefont {Liu},
  \citenamefont {Wang}, \citenamefont {Feng},\ and\ \citenamefont
  {Cao}}]{BaTiCrAsO}%
  \BibitemOpen
  \bibfield  {author} {\bibinfo {author} {\bibfnamefont {A.}~\bibnamefont
  {Ablimit}}, \bibinfo {author} {\bibfnamefont {Y.-L.}\ \bibnamefont {Sun}},
  \bibinfo {author} {\bibfnamefont {H.}~\bibnamefont {Jiang}}, \bibinfo
  {author} {\bibfnamefont {J.-K.}\ \bibnamefont {Bao}}, \bibinfo {author}
  {\bibfnamefont {H.-F.}\ \bibnamefont {Zhai}}, \bibinfo {author}
  {\bibfnamefont {Z.-T.}\ \bibnamefont {Tang}}, \bibinfo {author}
  {\bibfnamefont {Y.}~\bibnamefont {Liu}}, \bibinfo {author} {\bibfnamefont
  {Z.-C.}\ \bibnamefont {Wang}}, \bibinfo {author} {\bibfnamefont {C.-M.}\
  \bibnamefont {Feng}},\ and\ \bibinfo {author} {\bibfnamefont {G.-H.}\
  \bibnamefont {Cao}},\ }\bibfield  {title} {\bibinfo {title} {Synthesis,
  crystal structure and physical properties of a new oxypnictide
  {Ba$_2$Ti$_2$Cr$_2$As$_4$O} containing {[Ti$_2$As$_2$O]$^{2-}$} and
  {[Cr$_2$As$_2$]$^{2-}$} layers},\ }\href
  {https://doi.org/https://doi.org/10.1016/j.jallcom.2016.10.152} {\bibfield
  {journal} {\bibinfo  {journal} {J. Alloys Compd.}\ }\textbf {\bibinfo
  {volume} {694}},\ \bibinfo {pages} {1149} (\bibinfo {year}
  {2017})}\BibitemShut {NoStop}%
\bibitem [{\citenamefont {Lin}\ \emph {et~al.}(2022)\citenamefont {Lin},
  \citenamefont {Jiang}, \citenamefont {Li}, \citenamefont {Song},
  \citenamefont {Wu}, \citenamefont {Ren},\ and\ \citenamefont
  {Cao}}]{Sr2Cr2AsO3}%
  \BibitemOpen
  \bibfield  {author} {\bibinfo {author} {\bibfnamefont {Y.-Q.}\ \bibnamefont
  {Lin}}, \bibinfo {author} {\bibfnamefont {H.}~\bibnamefont {Jiang}}, \bibinfo
  {author} {\bibfnamefont {H.-X.}\ \bibnamefont {Li}}, \bibinfo {author}
  {\bibfnamefont {S.-J.}\ \bibnamefont {Song}}, \bibinfo {author}
  {\bibfnamefont {S.-Q.}\ \bibnamefont {Wu}}, \bibinfo {author} {\bibfnamefont
  {Z.}~\bibnamefont {Ren}},\ and\ \bibinfo {author} {\bibfnamefont {G.-H.}\
  \bibnamefont {Cao}},\ }\bibfield  {title} {\bibinfo {title} {Structural,
  electronic, and physical properties of a new layered {Cr}-based oxyarsenide
  {Sr$_2$Cr$_2$AsO$_3$}},\ }\href {https://doi.org/10.3390/ma15030802}
  {\bibfield  {journal} {\bibinfo  {journal} {Materials}\ }\textbf {\bibinfo
  {volume} {15}},\ \bibinfo {pages} {802} (\bibinfo {year} {2022})}\BibitemShut
  {NoStop}%
\bibitem [{\citenamefont {Chen}\ and\ \citenamefont
  {Wang}(2018)}]{CrReview_2019}%
  \BibitemOpen
  \bibfield  {author} {\bibinfo {author} {\bibfnamefont {R.~Y.}\ \bibnamefont
  {Chen}}\ and\ \bibinfo {author} {\bibfnamefont {N.~L.}\ \bibnamefont
  {Wang}},\ }\bibfield  {title} {\bibinfo {title} {Progress in {Cr}- and
  {Mn}-based superconductors: a key issues review},\ }\href
  {https://doi.org/10.1088/1361-6633/aaed0d} {\bibfield  {journal} {\bibinfo
  {journal} {Rep. Prog. Phys.}\ }\textbf {\bibinfo {volume} {82}},\ \bibinfo
  {pages} {012503} (\bibinfo {year} {2018})}\BibitemShut {NoStop}%
\bibitem [{\citenamefont {Wu}\ \emph {et~al.}(2014)\citenamefont {Wu},
  \citenamefont {Cheng}, \citenamefont {Matsubayashi}, \citenamefont {Kong},
  \citenamefont {Lin}, \citenamefont {Jin}, \citenamefont {Wang}, \citenamefont
  {Uwatoko},\ and\ \citenamefont {Luo}}]{CrAs_2014}%
  \BibitemOpen
  \bibfield  {author} {\bibinfo {author} {\bibfnamefont {W.}~\bibnamefont
  {Wu}}, \bibinfo {author} {\bibfnamefont {J.}~\bibnamefont {Cheng}}, \bibinfo
  {author} {\bibfnamefont {K.}~\bibnamefont {Matsubayashi}}, \bibinfo {author}
  {\bibfnamefont {P.}~\bibnamefont {Kong}}, \bibinfo {author} {\bibfnamefont
  {F.}~\bibnamefont {Lin}}, \bibinfo {author} {\bibfnamefont {C.}~\bibnamefont
  {Jin}}, \bibinfo {author} {\bibfnamefont {N.}~\bibnamefont {Wang}}, \bibinfo
  {author} {\bibfnamefont {Y.}~\bibnamefont {Uwatoko}},\ and\ \bibinfo {author}
  {\bibfnamefont {J.}~\bibnamefont {Luo}},\ }\bibfield  {title} {\bibinfo
  {title} {Superconductivity in the vicinity of antiferromagnetic order in
  {CrAs}},\ }\href {https://doi.org/10.1038/ncomms6508} {\bibfield  {journal}
  {\bibinfo  {journal} {Nat. Commun.}\ }\textbf {\bibinfo {volume} {5}},\
  \bibinfo {pages} {5508} (\bibinfo {year} {2014})}\BibitemShut {NoStop}%
\bibitem [{\citenamefont {Bao}\ \emph {et~al.}(2015)\citenamefont {Bao},
  \citenamefont {Liu}, \citenamefont {Ma}, \citenamefont {Meng}, \citenamefont
  {Tang}, \citenamefont {Sun}, \citenamefont {Zhai}, \citenamefont {Jiang},
  \citenamefont {Bai}, \citenamefont {Feng}, \citenamefont {Xu},\ and\
  \citenamefont {Cao}}]{K2Cr3As3_2015}%
  \BibitemOpen
  \bibfield  {author} {\bibinfo {author} {\bibfnamefont {J.-K.}\ \bibnamefont
  {Bao}}, \bibinfo {author} {\bibfnamefont {J.-Y.}\ \bibnamefont {Liu}},
  \bibinfo {author} {\bibfnamefont {C.-W.}\ \bibnamefont {Ma}}, \bibinfo
  {author} {\bibfnamefont {Z.-H.}\ \bibnamefont {Meng}}, \bibinfo {author}
  {\bibfnamefont {Z.-T.}\ \bibnamefont {Tang}}, \bibinfo {author}
  {\bibfnamefont {Y.-L.}\ \bibnamefont {Sun}}, \bibinfo {author} {\bibfnamefont
  {H.-F.}\ \bibnamefont {Zhai}}, \bibinfo {author} {\bibfnamefont
  {H.}~\bibnamefont {Jiang}}, \bibinfo {author} {\bibfnamefont
  {H.}~\bibnamefont {Bai}}, \bibinfo {author} {\bibfnamefont {C.-M.}\
  \bibnamefont {Feng}}, \bibinfo {author} {\bibfnamefont {Z.-A.}\ \bibnamefont
  {Xu}},\ and\ \bibinfo {author} {\bibfnamefont {G.-H.}\ \bibnamefont {Cao}},\
  }\bibfield  {title} {\bibinfo {title} {Superconductivity in
  {Quasi-One-Dimensional}
  {${\mathrm{K}}_{2}{\mathrm{Cr}}_{3}{\mathrm{As}}_{3}$} with {Significant}
  {Electron} {Correlations}},\ }\href
  {https://doi.org/10.1103/PhysRevX.5.011013} {\bibfield  {journal} {\bibinfo
  {journal} {Phys. Rev. X}\ }\textbf {\bibinfo {volume} {5}},\ \bibinfo {pages}
  {011013} (\bibinfo {year} {2015})}\BibitemShut {NoStop}%
\bibitem [{\citenamefont {Mu}\ \emph {et~al.}(2017)\citenamefont {Mu},
  \citenamefont {Ruan}, \citenamefont {Pan}, \citenamefont {Liu}, \citenamefont
  {Yu}, \citenamefont {Zhao}, \citenamefont {Chen},\ and\ \citenamefont
  {Ren}}]{KCr3As3_2017}%
  \BibitemOpen
  \bibfield  {author} {\bibinfo {author} {\bibfnamefont {Q.-G.}\ \bibnamefont
  {Mu}}, \bibinfo {author} {\bibfnamefont {B.-B.}\ \bibnamefont {Ruan}},
  \bibinfo {author} {\bibfnamefont {B.-J.}\ \bibnamefont {Pan}}, \bibinfo
  {author} {\bibfnamefont {T.}~\bibnamefont {Liu}}, \bibinfo {author}
  {\bibfnamefont {J.}~\bibnamefont {Yu}}, \bibinfo {author} {\bibfnamefont
  {K.}~\bibnamefont {Zhao}}, \bibinfo {author} {\bibfnamefont {G.-F.}\
  \bibnamefont {Chen}},\ and\ \bibinfo {author} {\bibfnamefont {Z.-A.}\
  \bibnamefont {Ren}},\ }\bibfield  {title} {\bibinfo {title}
  {Superconductivity at 5 {K} in quasi-one-dimensional {Cr}-based
  {${\mathrm{KCr}}_{3}{\mathrm{As}}_{3}$} single crystals},\ }\href
  {https://doi.org/10.1103/PhysRevB.96.140504} {\bibfield  {journal} {\bibinfo
  {journal} {Phys. Rev. B}\ }\textbf {\bibinfo {volume} {96}},\ \bibinfo
  {pages} {140504(R)} (\bibinfo {year} {2017})}\BibitemShut {NoStop}%
\bibitem [{\citenamefont {Xiang}\ \emph {et~al.}(2019)\citenamefont {Xiang},
  \citenamefont {Yu}, \citenamefont {Wu}, \citenamefont {Li}, \citenamefont
  {Shao}, \citenamefont {Tang}, \citenamefont {Bao},\ and\ \citenamefont
  {Cao}}]{K1331_2019}%
  \BibitemOpen
  \bibfield  {author} {\bibinfo {author} {\bibfnamefont {J.-J.}\ \bibnamefont
  {Xiang}}, \bibinfo {author} {\bibfnamefont {Y.-L.}\ \bibnamefont {Yu}},
  \bibinfo {author} {\bibfnamefont {S.-Q.}\ \bibnamefont {Wu}}, \bibinfo
  {author} {\bibfnamefont {B.-Z.}\ \bibnamefont {Li}}, \bibinfo {author}
  {\bibfnamefont {Y.-T.}\ \bibnamefont {Shao}}, \bibinfo {author}
  {\bibfnamefont {Z.-T.}\ \bibnamefont {Tang}}, \bibinfo {author}
  {\bibfnamefont {J.-K.}\ \bibnamefont {Bao}},\ and\ \bibinfo {author}
  {\bibfnamefont {G.-H.}\ \bibnamefont {Cao}},\ }\bibfield  {title} {\bibinfo
  {title} {Superconductivity induced by aging and annealing in
  {${\mathrm{K}}_{1\ensuremath{-}\ensuremath{\delta}}{\mathrm{Cr}}_{3}{\mathrm{As}}_{3}{\mathrm{H}}_{x}$}},\
  }\href {https://doi.org/10.1103/PhysRevMaterials.3.114802} {\bibfield
  {journal} {\bibinfo  {journal} {Phys. Rev. Mater.}\ }\textbf {\bibinfo
  {volume} {3}},\ \bibinfo {pages} {114802} (\bibinfo {year}
  {2019})}\BibitemShut {NoStop}%
\bibitem [{\citenamefont {Wu}\ \emph {et~al.}(2019)\citenamefont {Wu},
  \citenamefont {Liu}, \citenamefont {Li}, \citenamefont {Yu}, \citenamefont
  {Wu}, \citenamefont {Shao}, \citenamefont {Na}, \citenamefont {Li},
  \citenamefont {Huang}, \citenamefont {Xiang},\ and\ \citenamefont
  {Luo}}]{Pr3Cr10-xN11_2020}%
  \BibitemOpen
  \bibfield  {author} {\bibinfo {author} {\bibfnamefont {W.}~\bibnamefont
  {Wu}}, \bibinfo {author} {\bibfnamefont {K.}~\bibnamefont {Liu}}, \bibinfo
  {author} {\bibfnamefont {Y.}~\bibnamefont {Li}}, \bibinfo {author}
  {\bibfnamefont {Z.}~\bibnamefont {Yu}}, \bibinfo {author} {\bibfnamefont
  {D.}~\bibnamefont {Wu}}, \bibinfo {author} {\bibfnamefont {Y.}~\bibnamefont
  {Shao}}, \bibinfo {author} {\bibfnamefont {S.}~\bibnamefont {Na}}, \bibinfo
  {author} {\bibfnamefont {G.}~\bibnamefont {Li}}, \bibinfo {author}
  {\bibfnamefont {R.}~\bibnamefont {Huang}}, \bibinfo {author} {\bibfnamefont
  {T.}~\bibnamefont {Xiang}},\ and\ \bibinfo {author} {\bibfnamefont
  {J.}~\bibnamefont {Luo}},\ }\bibfield  {title} {\bibinfo {title}
  {{Superconductivity in chromium nitrides {Pr$_3$Cr$_{10-x}$N$_{11}$} with
  strong electron correlations}},\ }\href {https://doi.org/10.1093/nsr/nwz129}
  {\bibfield  {journal} {\bibinfo  {journal} {Natl. Sci. Rev.}\ }\textbf
  {\bibinfo {volume} {7}},\ \bibinfo {pages} {21} (\bibinfo {year}
  {2019})}\BibitemShut {NoStop}%
\bibitem [{\citenamefont {Edelmann}\ \emph {et~al.}(2017)\citenamefont
  {Edelmann}, \citenamefont {Sangiovanni}, \citenamefont {Capone},\ and\
  \citenamefont {de' Medici}}]{BaCr2As2_2017}%
  \BibitemOpen
  \bibfield  {author} {\bibinfo {author} {\bibfnamefont {M.}~\bibnamefont
  {Edelmann}}, \bibinfo {author} {\bibfnamefont {G.}~\bibnamefont
  {Sangiovanni}}, \bibinfo {author} {\bibfnamefont {M.}~\bibnamefont
  {Capone}},\ and\ \bibinfo {author} {\bibfnamefont {L.}~\bibnamefont {de'
  Medici}},\ }\bibfield  {title} {\bibinfo {title} {Chromium analogs of
  iron-based superconductors},\ }\href
  {https://doi.org/10.1103/PhysRevB.95.205118} {\bibfield  {journal} {\bibinfo
  {journal} {Phys. Rev. B}\ }\textbf {\bibinfo {volume} {95}},\ \bibinfo
  {pages} {205118} (\bibinfo {year} {2017})}\BibitemShut {NoStop}%
\bibitem [{\citenamefont {Pizarro}\ \emph {et~al.}(2017)\citenamefont
  {Pizarro}, \citenamefont {Calder\'on}, \citenamefont {Liu}, \citenamefont
  {Mu\~noz},\ and\ \citenamefont {Bascones}}]{LaCrAsO_2017_Pizarro}%
  \BibitemOpen
  \bibfield  {author} {\bibinfo {author} {\bibfnamefont {J.~M.}\ \bibnamefont
  {Pizarro}}, \bibinfo {author} {\bibfnamefont {M.~J.}\ \bibnamefont
  {Calder\'on}}, \bibinfo {author} {\bibfnamefont {J.}~\bibnamefont {Liu}},
  \bibinfo {author} {\bibfnamefont {M.~C.}\ \bibnamefont {Mu\~noz}},\ and\
  \bibinfo {author} {\bibfnamefont {E.}~\bibnamefont {Bascones}},\ }\bibfield
  {title} {\bibinfo {title} {Strong correlations and the search for
  high-${T}_{c}$ superconductivity in chromium pnictides and chalcogenides},\
  }\href {https://doi.org/10.1103/PhysRevB.95.075115} {\bibfield  {journal}
  {\bibinfo  {journal} {Phys. Rev. B}\ }\textbf {\bibinfo {volume} {95}},\
  \bibinfo {pages} {075115} (\bibinfo {year} {2017})}\BibitemShut {NoStop}%
\bibitem [{\citenamefont {Wang}\ \emph
  {et~al.}(2017{\natexlab{a}})\citenamefont {Wang}, \citenamefont {Gao},
  \citenamefont {Yang}, \citenamefont {Xiang},\ and\ \citenamefont
  {Wang}}]{LaCrAsO_2017_Wang}%
  \BibitemOpen
  \bibfield  {author} {\bibinfo {author} {\bibfnamefont {W.-S.}\ \bibnamefont
  {Wang}}, \bibinfo {author} {\bibfnamefont {M.}~\bibnamefont {Gao}}, \bibinfo
  {author} {\bibfnamefont {Y.}~\bibnamefont {Yang}}, \bibinfo {author}
  {\bibfnamefont {Y.-Y.}\ \bibnamefont {Xiang}},\ and\ \bibinfo {author}
  {\bibfnamefont {Q.-H.}\ \bibnamefont {Wang}},\ }\bibfield  {title} {\bibinfo
  {title} {Possible superconductivity in the electron-doped chromium pnictide
  {LaOCrAs}},\ }\href {https://doi.org/10.1103/PhysRevB.95.144507} {\bibfield
  {journal} {\bibinfo  {journal} {Phys. Rev. B}\ }\textbf {\bibinfo {volume}
  {95}},\ \bibinfo {pages} {144507} (\bibinfo {year}
  {2017}{\natexlab{a}})}\BibitemShut {NoStop}%
\bibitem [{\citenamefont {Ishida}\ and\ \citenamefont
  {Liebsch}(2010)}]{MottScenario_2011}%
  \BibitemOpen
  \bibfield  {author} {\bibinfo {author} {\bibfnamefont {H.}~\bibnamefont
  {Ishida}}\ and\ \bibinfo {author} {\bibfnamefont {A.}~\bibnamefont
  {Liebsch}},\ }\bibfield  {title} {\bibinfo {title} {{Fermi}-liquid,
  non-{Fermi}-liquid, and {Mott} phases in iron pnictides and cuprates},\
  }\href {https://doi.org/10.1103/PhysRevB.81.054513} {\bibfield  {journal}
  {\bibinfo  {journal} {Phys. Rev. B}\ }\textbf {\bibinfo {volume} {81}},\
  \bibinfo {pages} {054513} (\bibinfo {year} {2010})}\BibitemShut {NoStop}%
\bibitem [{\citenamefont {Park}\ \emph {et~al.}(2017)\citenamefont {Park},
  \citenamefont {Mizoguchi}, \citenamefont {Hiraka}, \citenamefont {Ikeda},
  \citenamefont {Otomo},\ and\ \citenamefont {Hosono}}]{La2Cr2As2OyHx_2017}%
  \BibitemOpen
  \bibfield  {author} {\bibinfo {author} {\bibfnamefont {S.-W.}\ \bibnamefont
  {Park}}, \bibinfo {author} {\bibfnamefont {H.}~\bibnamefont {Mizoguchi}},
  \bibinfo {author} {\bibfnamefont {H.}~\bibnamefont {Hiraka}}, \bibinfo
  {author} {\bibfnamefont {K.}~\bibnamefont {Ikeda}}, \bibinfo {author}
  {\bibfnamefont {T.}~\bibnamefont {Otomo}},\ and\ \bibinfo {author}
  {\bibfnamefont {H.}~\bibnamefont {Hosono}},\ }\bibfield  {title} {\bibinfo
  {title} {Transformation of the chromium coordination environment in {LaCrAsO}
  induced by hydride doping: Formation of {La$_2$Cr$_2$As$_2$O$_y$H$_x$}},\
  }\href {https://doi.org/10.1021/acs.inorgchem.7b02316} {\bibfield  {journal}
  {\bibinfo  {journal} {Inorg. Chem.}\ }\textbf {\bibinfo {volume} {56}},\
  \bibinfo {pages} {13642} (\bibinfo {year} {2017})}\BibitemShut {NoStop}%
\bibitem [{\citenamefont {Filsinger}\ \emph {et~al.}(2017)\citenamefont
  {Filsinger}, \citenamefont {Schnelle}, \citenamefont {Adler}, \citenamefont
  {Fecher}, \citenamefont {Reehuis}, \citenamefont {Hoser}, \citenamefont
  {Hoffmann}, \citenamefont {Werner}, \citenamefont {Greenblatt},\ and\
  \citenamefont {Felser}}]{BaCrFeAs_2017}%
  \BibitemOpen
  \bibfield  {author} {\bibinfo {author} {\bibfnamefont {K.~A.}\ \bibnamefont
  {Filsinger}}, \bibinfo {author} {\bibfnamefont {W.}~\bibnamefont {Schnelle}},
  \bibinfo {author} {\bibfnamefont {P.}~\bibnamefont {Adler}}, \bibinfo
  {author} {\bibfnamefont {G.~H.}\ \bibnamefont {Fecher}}, \bibinfo {author}
  {\bibfnamefont {M.}~\bibnamefont {Reehuis}}, \bibinfo {author} {\bibfnamefont
  {A.}~\bibnamefont {Hoser}}, \bibinfo {author} {\bibfnamefont {J.-U.}\
  \bibnamefont {Hoffmann}}, \bibinfo {author} {\bibfnamefont {P.}~\bibnamefont
  {Werner}}, \bibinfo {author} {\bibfnamefont {M.}~\bibnamefont {Greenblatt}},\
  and\ \bibinfo {author} {\bibfnamefont {C.}~\bibnamefont {Felser}},\
  }\bibfield  {title} {\bibinfo {title} {Antiferromagnetic structure and
  electronic properties of {${\mathrm{BaCr}}_{2}{\mathrm{As}}_{2}$} and
  {${\mathrm{BaCrFeAs}}_{2}$}},\ }\href
  {https://doi.org/10.1103/PhysRevB.95.184414} {\bibfield  {journal} {\bibinfo
  {journal} {Phys. Rev. B}\ }\textbf {\bibinfo {volume} {95}},\ \bibinfo
  {pages} {184414} (\bibinfo {year} {2017})}\BibitemShut {NoStop}%
\bibitem [{\citenamefont {Wang}\ \emph {et~al.}(2016)\citenamefont {Wang},
  \citenamefont {Wang}, \citenamefont {Mei}, \citenamefont {Li}, \citenamefont
  {Li}, \citenamefont {Tang}, \citenamefont {Liu}, \citenamefont {Zhang},
  \citenamefont {Zhai}, \citenamefont {Xu},\ and\ \citenamefont
  {Cao}}]{ThFeAsN}%
  \BibitemOpen
  \bibfield  {author} {\bibinfo {author} {\bibfnamefont {C.}~\bibnamefont
  {Wang}}, \bibinfo {author} {\bibfnamefont {Z.-C.}\ \bibnamefont {Wang}},
  \bibinfo {author} {\bibfnamefont {Y.-X.}\ \bibnamefont {Mei}}, \bibinfo
  {author} {\bibfnamefont {Y.-K.}\ \bibnamefont {Li}}, \bibinfo {author}
  {\bibfnamefont {L.}~\bibnamefont {Li}}, \bibinfo {author} {\bibfnamefont
  {Z.-T.}\ \bibnamefont {Tang}}, \bibinfo {author} {\bibfnamefont
  {Y.}~\bibnamefont {Liu}}, \bibinfo {author} {\bibfnamefont {P.}~\bibnamefont
  {Zhang}}, \bibinfo {author} {\bibfnamefont {H.-F.}\ \bibnamefont {Zhai}},
  \bibinfo {author} {\bibfnamefont {Z.-A.}\ \bibnamefont {Xu}},\ and\ \bibinfo
  {author} {\bibfnamefont {G.-H.}\ \bibnamefont {Cao}},\ }\bibfield  {title}
  {\bibinfo {title} {A new {ZrCuSiAs}-type superconductor: {ThFeAsN}},\ }\href
  {https://doi.org/10.1021/jacs.6b00236} {\bibfield  {journal} {\bibinfo
  {journal} {J. Am. Chem. Soc.}\ }\textbf {\bibinfo {volume} {138}},\ \bibinfo
  {pages} {2170} (\bibinfo {year} {2016})}\BibitemShut {NoStop}%
\bibitem [{\citenamefont {Li}\ \emph {et~al.}(2018)\citenamefont {Li},
  \citenamefont {Wang}, \citenamefont {Wang}, \citenamefont {Zhang},
  \citenamefont {Wang}, \citenamefont {Zhang}, \citenamefont {Sun},
  \citenamefont {Jing}, \citenamefont {Zhang}, \citenamefont {Tan},
  \citenamefont {Li}, \citenamefont {Feng}, \citenamefont {Mei}, \citenamefont
  {Wang},\ and\ \citenamefont {Cao}}]{ThFeAsN1-xOx}%
  \BibitemOpen
  \bibfield  {author} {\bibinfo {author} {\bibfnamefont {B.-Z.}\ \bibnamefont
  {Li}}, \bibinfo {author} {\bibfnamefont {Z.-C.}\ \bibnamefont {Wang}},
  \bibinfo {author} {\bibfnamefont {J.-L.}\ \bibnamefont {Wang}}, \bibinfo
  {author} {\bibfnamefont {F.-X.}\ \bibnamefont {Zhang}}, \bibinfo {author}
  {\bibfnamefont {D.-Z.}\ \bibnamefont {Wang}}, \bibinfo {author}
  {\bibfnamefont {F.-Y.}\ \bibnamefont {Zhang}}, \bibinfo {author}
  {\bibfnamefont {Y.-P.}\ \bibnamefont {Sun}}, \bibinfo {author} {\bibfnamefont
  {Q.}~\bibnamefont {Jing}}, \bibinfo {author} {\bibfnamefont {H.-F.}\
  \bibnamefont {Zhang}}, \bibinfo {author} {\bibfnamefont {S.-G.}\ \bibnamefont
  {Tan}}, \bibinfo {author} {\bibfnamefont {Y.-K.}\ \bibnamefont {Li}},
  \bibinfo {author} {\bibfnamefont {C.-M.}\ \bibnamefont {Feng}}, \bibinfo
  {author} {\bibfnamefont {Y.-X.}\ \bibnamefont {Mei}}, \bibinfo {author}
  {\bibfnamefont {C.}~\bibnamefont {Wang}},\ and\ \bibinfo {author}
  {\bibfnamefont {G.-H.}\ \bibnamefont {Cao}},\ }\bibfield  {title} {\bibinfo
  {title} {Peculiar phase diagram with isolated superconducting regions in
  {ThFeAsN$_{1-x}$O$_x$}},\ }\href {https://doi.org/10.1088/1361-648X/aac402}
  {\bibfield  {journal} {\bibinfo  {journal} {J. Phys.: Condens. Matter}\
  }\textbf {\bibinfo {volume} {30}},\ \bibinfo {pages} {255602} (\bibinfo
  {year} {2018})}\BibitemShut {NoStop}%
\bibitem [{\citenamefont {Shao}\ \emph {et~al.}(2019)\citenamefont {Shao},
  \citenamefont {Wang}, \citenamefont {Li}, \citenamefont {Wu}, \citenamefont
  {Wu}, \citenamefont {Ren}, \citenamefont {Qiu}, \citenamefont {Rao},
  \citenamefont {Wang},\ and\ \citenamefont {Cao}}]{BaTh}%
  \BibitemOpen
  \bibfield  {author} {\bibinfo {author} {\bibfnamefont {Y.-T.}\ \bibnamefont
  {Shao}}, \bibinfo {author} {\bibfnamefont {Z.-C.}\ \bibnamefont {Wang}},
  \bibinfo {author} {\bibfnamefont {B.-Z.}\ \bibnamefont {Li}}, \bibinfo
  {author} {\bibfnamefont {S.-Q.}\ \bibnamefont {Wu}}, \bibinfo {author}
  {\bibfnamefont {J.-F.}\ \bibnamefont {Wu}}, \bibinfo {author} {\bibfnamefont
  {Z.}~\bibnamefont {Ren}}, \bibinfo {author} {\bibfnamefont {S.-W.}\
  \bibnamefont {Qiu}}, \bibinfo {author} {\bibfnamefont {C.}~\bibnamefont
  {Rao}}, \bibinfo {author} {\bibfnamefont {C.}~\bibnamefont {Wang}},\ and\
  \bibinfo {author} {\bibfnamefont {G.-H.}\ \bibnamefont {Cao}},\ }\bibfield
  {title} {\bibinfo {title} {{BaTh$_2$Fe$_4$As$_4$(N$_{0.7}$O$_{0.3}$)$_2$}: An
  iron-based superconductor stabilized by inter-block-layer charge transfer},\
  }\href {https://doi.org/10.1007/s40843-019-9438-7} {\bibfield  {journal}
  {\bibinfo  {journal} {Sci. China Mater.}\ }\textbf {\bibinfo {volume} {62}},\
  \bibinfo {pages} {1357} (\bibinfo {year} {2019})}\BibitemShut {NoStop}%
\bibitem [{\citenamefont {Wang}\ \emph
  {et~al.}(2017{\natexlab{b}})\citenamefont {Wang}, \citenamefont {Shao},
  \citenamefont {Wang}, \citenamefont {Wang}, \citenamefont {Xu},\ and\
  \citenamefont {Cao}}]{ThNiAsN}%
  \BibitemOpen
  \bibfield  {author} {\bibinfo {author} {\bibfnamefont {Z.-C.}\ \bibnamefont
  {Wang}}, \bibinfo {author} {\bibfnamefont {Y.-T.}\ \bibnamefont {Shao}},
  \bibinfo {author} {\bibfnamefont {C.}~\bibnamefont {Wang}}, \bibinfo {author}
  {\bibfnamefont {Z.}~\bibnamefont {Wang}}, \bibinfo {author} {\bibfnamefont
  {Z.-A.}\ \bibnamefont {Xu}},\ and\ \bibinfo {author} {\bibfnamefont {G.-H.}\
  \bibnamefont {Cao}},\ }\bibfield  {title} {\bibinfo {title} {Enhanced
  superconductivity in {ThNiAsN}},\ }\href
  {https://doi.org/10.1209/0295-5075/118/57004} {\bibfield  {journal} {\bibinfo
   {journal} {EPL}\ }\textbf {\bibinfo {volume} {118}},\ \bibinfo {pages}
  {57004} (\bibinfo {year} {2017}{\natexlab{b}})}\BibitemShut {NoStop}%
\bibitem [{\citenamefont {Zhang}\ \emph {et~al.}(2020)\citenamefont {Zhang},
  \citenamefont {Li}, \citenamefont {Ren}, \citenamefont {Mao}, \citenamefont
  {Xia}, \citenamefont {Hu}, \citenamefont {Liu}, \citenamefont {Wang},
  \citenamefont {Shao}, \citenamefont {Feng}, \citenamefont {Tan},
  \citenamefont {Sun}, \citenamefont {Ren}, \citenamefont {Jing}, \citenamefont
  {Liu}, \citenamefont {Luo}, \citenamefont {Ma}, \citenamefont {Mei},
  \citenamefont {Wang},\ and\ \citenamefont {Cao}}]{ThMnPN}%
  \BibitemOpen
  \bibfield  {author} {\bibinfo {author} {\bibfnamefont {F.}~\bibnamefont
  {Zhang}}, \bibinfo {author} {\bibfnamefont {B.}~\bibnamefont {Li}}, \bibinfo
  {author} {\bibfnamefont {Q.}~\bibnamefont {Ren}}, \bibinfo {author}
  {\bibfnamefont {H.}~\bibnamefont {Mao}}, \bibinfo {author} {\bibfnamefont
  {Y.}~\bibnamefont {Xia}}, \bibinfo {author} {\bibfnamefont {B.}~\bibnamefont
  {Hu}}, \bibinfo {author} {\bibfnamefont {Z.}~\bibnamefont {Liu}}, \bibinfo
  {author} {\bibfnamefont {Z.}~\bibnamefont {Wang}}, \bibinfo {author}
  {\bibfnamefont {Y.}~\bibnamefont {Shao}}, \bibinfo {author} {\bibfnamefont
  {Z.}~\bibnamefont {Feng}}, \bibinfo {author} {\bibfnamefont {S.}~\bibnamefont
  {Tan}}, \bibinfo {author} {\bibfnamefont {Y.}~\bibnamefont {Sun}}, \bibinfo
  {author} {\bibfnamefont {Z.}~\bibnamefont {Ren}}, \bibinfo {author}
  {\bibfnamefont {Q.}~\bibnamefont {Jing}}, \bibinfo {author} {\bibfnamefont
  {B.}~\bibnamefont {Liu}}, \bibinfo {author} {\bibfnamefont {H.}~\bibnamefont
  {Luo}}, \bibinfo {author} {\bibfnamefont {J.}~\bibnamefont {Ma}}, \bibinfo
  {author} {\bibfnamefont {Y.}~\bibnamefont {Mei}}, \bibinfo {author}
  {\bibfnamefont {C.}~\bibnamefont {Wang}},\ and\ \bibinfo {author}
  {\bibfnamefont {G.-H.}\ \bibnamefont {Cao}},\ }\bibfield  {title} {\bibinfo
  {title} {{ThMnPnN} ({Pn} = {P}, {As}): Synthesis, structure, and chemical
  pressure effects},\ }\href {https://doi.org/10.1021/acs.inorgchem.9b03294}
  {\bibfield  {journal} {\bibinfo  {journal} {Inorg. Chem.}\ }\textbf {\bibinfo
  {volume} {59}},\ \bibinfo {pages} {2937} (\bibinfo {year}
  {2020})}\BibitemShut {NoStop}%
\bibitem [{\citenamefont {Rodr\'{i}guez-Carvajal}(1993)}]{fullprof}%
  \BibitemOpen
  \bibfield  {author} {\bibinfo {author} {\bibfnamefont {J.}~\bibnamefont
  {Rodr\'{i}guez-Carvajal}},\ }\bibfield  {title} {\bibinfo {title} {Recent
  advances in magnetic structure determination by neutron powder diffraction},\
  }\href {https://doi.org/10.1016/0921-4526(93)90108-I} {\bibfield  {journal}
  {\bibinfo  {journal} {Phys. B: Condens. Matter}\ }\textbf {\bibinfo {volume}
  {192}},\ \bibinfo {pages} {55} (\bibinfo {year} {1993})}\BibitemShut
  {NoStop}%
\bibitem [{\citenamefont {Kresse}\ and\ \citenamefont
  {Furthm\"uller}(1996)}]{VASP}%
  \BibitemOpen
  \bibfield  {author} {\bibinfo {author} {\bibfnamefont {G.}~\bibnamefont
  {Kresse}}\ and\ \bibinfo {author} {\bibfnamefont {J.}~\bibnamefont
  {Furthm\"uller}},\ }\bibfield  {title} {\bibinfo {title} {Efficient iterative
  schemes for ab initio total-energy calculations using a plane-wave basis
  set},\ }\href {https://doi.org/10.1103/PhysRevB.54.11169} {\bibfield
  {journal} {\bibinfo  {journal} {Phys. Rev. B}\ }\textbf {\bibinfo {volume}
  {54}},\ \bibinfo {pages} {11169} (\bibinfo {year} {1996})}\BibitemShut
  {NoStop}%
\bibitem [{\citenamefont {Ueda}\ \emph {et~al.}(2004)\citenamefont {Ueda},
  \citenamefont {Hosono},\ and\ \citenamefont {Hamada}}]{La_4f}%
  \BibitemOpen
  \bibfield  {author} {\bibinfo {author} {\bibfnamefont {K.}~\bibnamefont
  {Ueda}}, \bibinfo {author} {\bibfnamefont {H.}~\bibnamefont {Hosono}},\ and\
  \bibinfo {author} {\bibfnamefont {N.}~\bibnamefont {Hamada}},\ }\bibfield
  {title} {\bibinfo {title} {Energy band structure of {LaCuOCh} ({Ch = S, Se
  and Te}) calculated by the full-potential linearized augmented plane-wave
  method},\ }\href {https://doi.org/10.1088/0953-8984/16/28/036} {\bibfield
  {journal} {\bibinfo  {journal} {J. Phys.: Condens. Matter}\ }\textbf
  {\bibinfo {volume} {16}},\ \bibinfo {pages} {5179} (\bibinfo {year}
  {2004})}\BibitemShut {NoStop}%
\bibitem [{sup()}]{suppmatt}%
  \BibitemOpen
  \href {https://journals.aps.org} {}\bibinfo {note} {See the Supplemental
  Material for X-ray diffraction for ThCrAsN$_{0.9}$O$_{0.1}$, crystallographic
  data of ThFeAsN$_{0.9}$O$_{0.1}$ at 300 K, magnetic susceptibility for
  ThCrAsN$_{1-x}$O$_x$ ($x\geq0.1$), calculated magnetic energies and Cr spin
  moments for ThCrAsO, and band structures of ThCrAsN and ThCrAsO.}\BibitemShut
  {Stop}%
\bibitem [{\citenamefont {Wang}\ \emph {et~al.}(2020)\citenamefont {Wang},
  \citenamefont {Jiao}, \citenamefont {Wang}, \citenamefont {Zhu},
  \citenamefont {Cai}, \citenamefont {Yang}, \citenamefont {Song},
  \citenamefont {Zhang}, \citenamefont {Li}, \citenamefont {Li}, \citenamefont
  {Hu}, \citenamefont {Li}, \citenamefont {Li}, \citenamefont {Tan},
  \citenamefont {Mei}, \citenamefont {Jing}, \citenamefont {Wang},
  \citenamefont {Liu},\ and\ \citenamefont {Qian}}]{ThCoAsN}%
  \BibitemOpen
  \bibfield  {author} {\bibinfo {author} {\bibfnamefont {J.}~\bibnamefont
  {Wang}}, \bibinfo {author} {\bibfnamefont {F.}~\bibnamefont {Jiao}}, \bibinfo
  {author} {\bibfnamefont {X.}~\bibnamefont {Wang}}, \bibinfo {author}
  {\bibfnamefont {S.}~\bibnamefont {Zhu}}, \bibinfo {author} {\bibfnamefont
  {L.}~\bibnamefont {Cai}}, \bibinfo {author} {\bibfnamefont {C.}~\bibnamefont
  {Yang}}, \bibinfo {author} {\bibfnamefont {P.}~\bibnamefont {Song}}, \bibinfo
  {author} {\bibfnamefont {F.}~\bibnamefont {Zhang}}, \bibinfo {author}
  {\bibfnamefont {B.}~\bibnamefont {Li}}, \bibinfo {author} {\bibfnamefont
  {Y.}~\bibnamefont {Li}}, \bibinfo {author} {\bibfnamefont {J.}~\bibnamefont
  {Hu}}, \bibinfo {author} {\bibfnamefont {S.}~\bibnamefont {Li}}, \bibinfo
  {author} {\bibfnamefont {Y.}~\bibnamefont {Li}}, \bibinfo {author}
  {\bibfnamefont {S.}~\bibnamefont {Tan}}, \bibinfo {author} {\bibfnamefont
  {Y.}~\bibnamefont {Mei}}, \bibinfo {author} {\bibfnamefont {Q.}~\bibnamefont
  {Jing}}, \bibinfo {author} {\bibfnamefont {C.}~\bibnamefont {Wang}}, \bibinfo
  {author} {\bibfnamefont {B.}~\bibnamefont {Liu}},\ and\ \bibinfo {author}
  {\bibfnamefont {D.}~\bibnamefont {Qian}},\ }\bibfield  {title} {\bibinfo
  {title} {The phase transition of {ThFe$_{1-x}$Co$_x$AsN} from superconductor
  to metallic paramagnet},\ }\href
  {https://doi.org/10.1209/0295-5075/130/67003} {\bibfield  {journal} {\bibinfo
   {journal} {EPL}\ }\textbf {\bibinfo {volume} {130}},\ \bibinfo {pages}
  {67003} (\bibinfo {year} {2020})}\BibitemShut {NoStop}%
\bibitem [{\citenamefont {Fisher}(1962)}]{T_N}%
  \BibitemOpen
  \bibfield  {author} {\bibinfo {author} {\bibfnamefont {M.~E.}\ \bibnamefont
  {Fisher}},\ }\bibfield  {title} {\bibinfo {title} {Relation between the
  specific heat and susceptibility of an antiferromagnet},\ }\href
  {https://doi.org/10.1080/14786436208213705} {\bibfield  {journal} {\bibinfo
  {journal} {Philos. Mag.}\ }\textbf {\bibinfo {volume} {7}},\ \bibinfo {pages}
  {1731} (\bibinfo {year} {1962})}\BibitemShut {NoStop}%
\bibitem [{\citenamefont {de~Jongh}\ and\ \citenamefont
  {Miedema}(1974)}]{T_N_2}%
  \BibitemOpen
  \bibfield  {author} {\bibinfo {author} {\bibfnamefont {L.}~\bibnamefont
  {de~Jongh}}\ and\ \bibinfo {author} {\bibfnamefont {A.}~\bibnamefont
  {Miedema}},\ }\bibfield  {title} {\bibinfo {title} {Experiments on simple
  magnetic model systems},\ }\href {https://doi.org/10.1080/00018739700101558}
  {\bibfield  {journal} {\bibinfo  {journal} {Adv. Phys.}\ }\textbf {\bibinfo
  {volume} {23}},\ \bibinfo {pages} {1} (\bibinfo {year} {1974})}\BibitemShut
  {NoStop}%
\bibitem [{\citenamefont {Nandi}\ \emph {et~al.}(2016)\citenamefont {Nandi},
  \citenamefont {Xiao}, \citenamefont {Qureshi}, \citenamefont {Paramanik},
  \citenamefont {Jin}, \citenamefont {Su}, \citenamefont {Ouladdiaf},
  \citenamefont {Hossain},\ and\ \citenamefont {Br\"uckel}}]{EuCr2As2_2016}%
  \BibitemOpen
  \bibfield  {author} {\bibinfo {author} {\bibfnamefont {S.}~\bibnamefont
  {Nandi}}, \bibinfo {author} {\bibfnamefont {Y.}~\bibnamefont {Xiao}},
  \bibinfo {author} {\bibfnamefont {N.}~\bibnamefont {Qureshi}}, \bibinfo
  {author} {\bibfnamefont {U.~B.}\ \bibnamefont {Paramanik}}, \bibinfo {author}
  {\bibfnamefont {W.~T.}\ \bibnamefont {Jin}}, \bibinfo {author} {\bibfnamefont
  {Y.}~\bibnamefont {Su}}, \bibinfo {author} {\bibfnamefont {B.}~\bibnamefont
  {Ouladdiaf}}, \bibinfo {author} {\bibfnamefont {Z.}~\bibnamefont {Hossain}},\
  and\ \bibinfo {author} {\bibfnamefont {T.}~\bibnamefont {Br\"uckel}},\
  }\bibfield  {title} {\bibinfo {title} {Magnetic structures of the {Eu} and
  {Cr} moments in {${\mathrm{EuCr}}_{2}{\mathrm{As}}_{2}$}: Neutron diffraction
  study},\ }\href {https://doi.org/10.1103/PhysRevB.94.094411} {\bibfield
  {journal} {\bibinfo  {journal} {Phys. Rev. B}\ }\textbf {\bibinfo {volume}
  {94}},\ \bibinfo {pages} {094411} (\bibinfo {year} {2016})}\BibitemShut
  {NoStop}%
\bibitem [{\citenamefont {Liu}\ \emph {et~al.}(2018)\citenamefont {Liu},
  \citenamefont {Wang}, \citenamefont {Sheng}, \citenamefont {Ye},
  \citenamefont {Taddei}, \citenamefont {Fernandez-Baca}, \citenamefont {Luo},
  \citenamefont {Sun}, \citenamefont {Wang}, \citenamefont {Jiang},
  \citenamefont {Cao},\ and\ \citenamefont {Bao}}]{Sr2Cr3As3O2_2018}%
  \BibitemOpen
  \bibfield  {author} {\bibinfo {author} {\bibfnamefont {J.}~\bibnamefont
  {Liu}}, \bibinfo {author} {\bibfnamefont {J.}~\bibnamefont {Wang}}, \bibinfo
  {author} {\bibfnamefont {J.}~\bibnamefont {Sheng}}, \bibinfo {author}
  {\bibfnamefont {F.}~\bibnamefont {Ye}}, \bibinfo {author} {\bibfnamefont
  {K.~M.}\ \bibnamefont {Taddei}}, \bibinfo {author} {\bibfnamefont {J.~A.}\
  \bibnamefont {Fernandez-Baca}}, \bibinfo {author} {\bibfnamefont
  {W.}~\bibnamefont {Luo}}, \bibinfo {author} {\bibfnamefont {G.-A.}\
  \bibnamefont {Sun}}, \bibinfo {author} {\bibfnamefont {Z.-C.}\ \bibnamefont
  {Wang}}, \bibinfo {author} {\bibfnamefont {H.}~\bibnamefont {Jiang}},
  \bibinfo {author} {\bibfnamefont {G.-H.}\ \bibnamefont {Cao}},\ and\ \bibinfo
  {author} {\bibfnamefont {W.}~\bibnamefont {Bao}},\ }\bibfield  {title}
  {\bibinfo {title} {Neutron diffraction study on magnetic structures and
  transitions in
  {${\mathrm{Sr}}_{2}{\mathrm{Cr}}_{3}{\mathrm{As}}_{2}{\mathrm{O}}_{2}$}},\
  }\href {https://doi.org/10.1103/PhysRevB.98.134416} {\bibfield  {journal}
  {\bibinfo  {journal} {Phys. Rev. B}\ }\textbf {\bibinfo {volume} {98}},\
  \bibinfo {pages} {134416} (\bibinfo {year} {2018})}\BibitemShut {NoStop}%
\bibitem [{\citenamefont {Jishi}\ \emph {et~al.}(2019)\citenamefont {Jishi},
  \citenamefont {Rodriguez}, \citenamefont {Haugan},\ and\ \citenamefont
  {Susner}}]{BaCr2P2_2019}%
  \BibitemOpen
  \bibfield  {author} {\bibinfo {author} {\bibfnamefont {R.~A.}\ \bibnamefont
  {Jishi}}, \bibinfo {author} {\bibfnamefont {J.~P.}\ \bibnamefont
  {Rodriguez}}, \bibinfo {author} {\bibfnamefont {T.~J.}\ \bibnamefont
  {Haugan}},\ and\ \bibinfo {author} {\bibfnamefont {M.~A.}\ \bibnamefont
  {Susner}},\ }\bibfield  {title} {\bibinfo {title} {Prediction of
  antiferromagnetism in barium chromium phosphide confirmed after synthesis},\
  }\href {https://doi.org/10.1088/1361-648X/ab45dc} {\bibfield  {journal}
  {\bibinfo  {journal} {J. Phys.: Condens. Matter}\ }\textbf {\bibinfo {volume}
  {32}},\ \bibinfo {pages} {025502} (\bibinfo {year} {2019})}\BibitemShut
  {NoStop}%
\bibitem [{\citenamefont {Johnston}\ \emph {et~al.}(2011)\citenamefont
  {Johnston}, \citenamefont {McQueeney}, \citenamefont {Lake}, \citenamefont
  {Honecker}, \citenamefont {Zhitomirsky}, \citenamefont {Nath}, \citenamefont
  {Furukawa}, \citenamefont {Antropov},\ and\ \citenamefont
  {Singh}}]{Heisenberg_model}%
  \BibitemOpen
  \bibfield  {author} {\bibinfo {author} {\bibfnamefont {D.~C.}\ \bibnamefont
  {Johnston}}, \bibinfo {author} {\bibfnamefont {R.~J.}\ \bibnamefont
  {McQueeney}}, \bibinfo {author} {\bibfnamefont {B.}~\bibnamefont {Lake}},
  \bibinfo {author} {\bibfnamefont {A.}~\bibnamefont {Honecker}}, \bibinfo
  {author} {\bibfnamefont {M.~E.}\ \bibnamefont {Zhitomirsky}}, \bibinfo
  {author} {\bibfnamefont {R.}~\bibnamefont {Nath}}, \bibinfo {author}
  {\bibfnamefont {Y.}~\bibnamefont {Furukawa}}, \bibinfo {author}
  {\bibfnamefont {V.~P.}\ \bibnamefont {Antropov}},\ and\ \bibinfo {author}
  {\bibfnamefont {Y.}~\bibnamefont {Singh}},\ }\bibfield  {title} {\bibinfo
  {title} {Magnetic exchange interactions in
  {${\mathrm{BaMn}}_{2}{\mathrm{As}}_{2}$}: A case study of the
  ${J}_{1}$-${J}_{2}$-${J}_{c}$ heisenberg model},\ }\href
  {https://doi.org/10.1103/PhysRevB.84.094445} {\bibfield  {journal} {\bibinfo
  {journal} {Phys. Rev. B}\ }\textbf {\bibinfo {volume} {84}},\ \bibinfo
  {pages} {094445} (\bibinfo {year} {2011})}\BibitemShut {NoStop}%
\bibitem [{\citenamefont {Ma}\ \emph {et~al.}(2008)\citenamefont {Ma},
  \citenamefont {Lu},\ and\ \citenamefont {Xiang}}]{LaFeAsO_J}%
  \BibitemOpen
  \bibfield  {author} {\bibinfo {author} {\bibfnamefont {F.}~\bibnamefont
  {Ma}}, \bibinfo {author} {\bibfnamefont {Z.-Y.}\ \bibnamefont {Lu}},\ and\
  \bibinfo {author} {\bibfnamefont {T.}~\bibnamefont {Xiang}},\ }\bibfield
  {title} {\bibinfo {title} {Arsenic-bridged antiferromagnetic superexchange
  interactions in {LaFeAsO}},\ }\href
  {https://doi.org/10.1103/PhysRevB.78.224517} {\bibfield  {journal} {\bibinfo
  {journal} {Phys. Rev. B}\ }\textbf {\bibinfo {volume} {78}},\ \bibinfo
  {pages} {224517} (\bibinfo {year} {2008})}\BibitemShut {NoStop}%
\bibitem [{\citenamefont {Yu}\ and\ \citenamefont {Li}(2012)}]{J2_PRB}%
  \BibitemOpen
  \bibfield  {author} {\bibinfo {author} {\bibfnamefont {S.-L.}\ \bibnamefont
  {Yu}}\ and\ \bibinfo {author} {\bibfnamefont {J.-X.}\ \bibnamefont {Li}},\
  }\bibfield  {title} {\bibinfo {title} {Chiral superconducting phase and
  chiral spin-density-wave phase in a hubbard model on the kagome lattice},\
  }\href {https://doi.org/10.1103/PhysRevB.85.144402} {\bibfield  {journal}
  {\bibinfo  {journal} {Phys. Rev. B}\ }\textbf {\bibinfo {volume} {85}},\
  \bibinfo {pages} {144402} (\bibinfo {year} {2012})}\BibitemShut {NoStop}%
\bibitem [{\citenamefont {Hu}\ and\ \citenamefont {Ding}(2012)}]{J2_SR}%
  \BibitemOpen
  \bibfield  {author} {\bibinfo {author} {\bibfnamefont {J.}~\bibnamefont
  {Hu}}\ and\ \bibinfo {author} {\bibfnamefont {H.}~\bibnamefont {Ding}},\
  }\bibfield  {title} {\bibinfo {title} {Local antiferromagnetic exchange and
  collaborative fermi surface as key ingredients of high temperature
  superconductors},\ }\href {https://doi.org/10.1038/srep00381} {\bibfield
  {journal} {\bibinfo  {journal} {Sci. Rep.}\ }\textbf {\bibinfo {volume}
  {2}},\ \bibinfo {pages} {381} (\bibinfo {year} {2012})}\BibitemShut {NoStop}%
\bibitem [{\citenamefont {Hu}\ \emph {et~al.}(2015)\citenamefont {Hu},
  \citenamefont {Le},\ and\ \citenamefont {Wu}}]{HuJP_2015}%
  \BibitemOpen
  \bibfield  {author} {\bibinfo {author} {\bibfnamefont {J.}~\bibnamefont
  {Hu}}, \bibinfo {author} {\bibfnamefont {C.}~\bibnamefont {Le}},\ and\
  \bibinfo {author} {\bibfnamefont {X.}~\bibnamefont {Wu}},\ }\bibfield
  {title} {\bibinfo {title} {Predicting unconventional high-temperature
  superconductors in trigonal bipyramidal coordinations},\ }\href
  {https://doi.org/10.1103/PhysRevX.5.041012} {\bibfield  {journal} {\bibinfo
  {journal} {Phys. Rev. X}\ }\textbf {\bibinfo {volume} {5}},\ \bibinfo {pages}
  {041012} (\bibinfo {year} {2015})}\BibitemShut {NoStop}%
\bibitem [{\citenamefont {Richard}\ \emph {et~al.}(2017)\citenamefont
  {Richard}, \citenamefont {van Roekeghem}, \citenamefont {Lv}, \citenamefont
  {Qian}, \citenamefont {Kim}, \citenamefont {Hoesch}, \citenamefont {Hu},
  \citenamefont {Sefat}, \citenamefont {Biermann},\ and\ \citenamefont
  {Ding}}]{BaCr2As2_ARPES}%
  \BibitemOpen
  \bibfield  {author} {\bibinfo {author} {\bibfnamefont {P.}~\bibnamefont
  {Richard}}, \bibinfo {author} {\bibfnamefont {A.}~\bibnamefont {van
  Roekeghem}}, \bibinfo {author} {\bibfnamefont {B.~Q.}\ \bibnamefont {Lv}},
  \bibinfo {author} {\bibfnamefont {T.}~\bibnamefont {Qian}}, \bibinfo {author}
  {\bibfnamefont {T.~K.}\ \bibnamefont {Kim}}, \bibinfo {author} {\bibfnamefont
  {M.}~\bibnamefont {Hoesch}}, \bibinfo {author} {\bibfnamefont {J.-P.}\
  \bibnamefont {Hu}}, \bibinfo {author} {\bibfnamefont {A.~S.}\ \bibnamefont
  {Sefat}}, \bibinfo {author} {\bibfnamefont {S.}~\bibnamefont {Biermann}},\
  and\ \bibinfo {author} {\bibfnamefont {H.}~\bibnamefont {Ding}},\ }\bibfield
  {title} {\bibinfo {title} {Is {${\mathrm{BaCr}}_{2}{\mathrm{As}}_{2}$}
  symmetrical to {${\mathrm{BaFe}}_{2}{\mathrm{As}}_{2}$} with respect to half
  $3d$ shell filling?},\ }\href {https://doi.org/10.1103/PhysRevB.95.184516}
  {\bibfield  {journal} {\bibinfo  {journal} {Phys. Rev. B}\ }\textbf {\bibinfo
  {volume} {95}},\ \bibinfo {pages} {184516} (\bibinfo {year}
  {2017})}\BibitemShut {NoStop}%
\end{thebibliography}%


\begin{thebibliography}{0}%
\makeatletter
\providecommand \@ifxundefined [1]{%
 \@ifx{#1\undefined}
}%
\providecommand \@ifnum [1]{%
 \ifnum #1\expandafter \@firstoftwo
 \else \expandafter \@secondoftwo
 \fi
}%
\providecommand \@ifx [1]{%
 \ifx #1\expandafter \@firstoftwo
 \else \expandafter \@secondoftwo
 \fi
}%
\providecommand \natexlab [1]{#1}%
\providecommand \enquote  [1]{``#1''}%
\providecommand \bibnamefont  [1]{#1}%
\providecommand \bibfnamefont [1]{#1}%
\providecommand \citenamefont [1]{#1}%
\providecommand \href@noop [0]{\@secondoftwo}%
\providecommand \href [0]{\begingroup \@sanitize@url \@href}%
\providecommand \@href[1]{\@@startlink{#1}\@@href}%
\providecommand \@@href[1]{\endgroup#1\@@endlink}%
\providecommand \@sanitize@url [0]{\catcode `\\12\catcode `\$12\catcode
  `\&12\catcode `\#12\catcode `\^12\catcode `\_12\catcode `\%12\relax}%
\providecommand \@@startlink[1]{}%
\providecommand \@@endlink[0]{}%
\providecommand \url  [0]{\begingroup\@sanitize@url \@url }%
\providecommand \@url [1]{\endgroup\@href {#1}{\urlprefix }}%
\providecommand \urlprefix  [0]{URL }%
\providecommand \Eprint [0]{\href }%
\providecommand \doibase [0]{https://doi.org/}%
\providecommand \selectlanguage [0]{\@gobble}%
\providecommand \bibinfo  [0]{\@secondoftwo}%
\providecommand \bibfield  [0]{\@secondoftwo}%
\providecommand \translation [1]{[#1]}%
\providecommand \BibitemOpen [0]{}%
\providecommand \bibitemStop [0]{}%
\providecommand \bibitemNoStop [0]{.\EOS\space}%
\providecommand \EOS [0]{\spacefactor3000\relax}%
\providecommand \BibitemShut  [1]{\csname bibitem#1\endcsname}%
\let\auto@bib@innerbib\@empty
\end{thebibliography}%

\end{document}



\title{Supplemental Materials: Absence of superconductivity in electron-doped chromium pnictides ThCrAsN$_{1-x}$O$_x$}

\author{Zhi-Cheng Wang}
\email{wzc@seu.edu.cn}
\affiliation{Key Laboratory of Quantum Materials and Devices of Ministry of Education, School of Physics, Southeast University, Nanjing 211189, China}
\author{Ye-Ting Shao}
\affiliation{School of Physics, Interdisciplinary Center for Quantum Information and State Key Laboratory of Silicon and Advanced Semiconductor Materials, Zhejiang University, Hangzhou 310058, China}
\author{Yi-Qiang Lin}
\affiliation{School of Physics, Interdisciplinary Center for Quantum Information and State Key Laboratory of Silicon and Advanced Semiconductor Materials, Zhejiang University, Hangzhou 310058, China}
\author{Shi-Jie Song}
\affiliation{School of Physics, Interdisciplinary Center for Quantum Information and State Key Laboratory of Silicon and Advanced Semiconductor Materials, Zhejiang University, Hangzhou 310058, China}
\author{Bai-Zhuo Li}
\affiliation{School of Physics, Interdisciplinary Center for Quantum Information and State Key Laboratory of Silicon and Advanced Semiconductor Materials, Zhejiang University, Hangzhou 310058, China}
\author{Er-Jian Cheng}
\affiliation{State Key Laboratory of Surface Physics and Department of Physics, Fudan University, Shanghai 200438, China}
\author{Shi-Yan Li}
\affiliation{State Key Laboratory of Surface Physics and Department of Physics, Fudan University, Shanghai 200438, China}
\affiliation{Collaborative Innovation Centre of Advanced Microstructures, Nanjing University, Nanjing 210093, China}
\affiliation{Shanghai Research Center for Quantum Sciences, Shanghai 201315, China}
\author{Qin-Qing Zhu}
\affiliation{School of Science, Westlake University, Hangzhou 310024, China}
\affiliation{Institute of Natural Sciences, Westlake Institute for Advanced Study, Hangzhou 310024, China}
\author{Zhi Ren}
\affiliation{School of Science, Westlake University, Hangzhou 310024, China}
\affiliation{Institute of Natural Sciences, Westlake Institute for Advanced Study, Hangzhou 310024, China}
\author{Guang-Han Cao}	
\email{ghcao@zju.edu.cn}
\affiliation{School of Physics, Interdisciplinary Center for Quantum Information and State Key Laboratory of Silicon and Advanced Semiconductor Materials, Zhejiang University, Hangzhou 310058, China}
\affiliation{Collaborative Innovation Centre of Advanced Microstructures, Nanjing University, Nanjing 210093, China}


\date{\today}

\maketitle

\begin{center}
	\textbf{Content}
	\begin{enumerate}[label=\Alph*.]
		\item Estimation of the ratio between the 1111 phase and ThO$_2$
		\item Figure S1: X-ray diffraction for ThCrAsN$_{0.9}$O$_{0.1}$
		\item Table S1: Crystallographic data of ThFeAsN$_{0.9}$O$_{0.1}$ at 300 K
		\item Figure S2: Magnetic susceptibility for ThCrAsN$_{1-x}$O$_x$
		\item Table S2: Calculated magnetic energies and Cr spin moments
		\item Figure S3: Band structures of ThCrAsN and ThCrAsO
	\end{enumerate}
\end{center}
\clearpage

\subsection{Estimation of the ratio between the 1111 phase and ThO$_2$}
We can further estimate the ratio of 1111 phase and ThO$_2$. Let’s take the pattern of ThCrAsN$_{0.1}$O$_{0.9}$ as an example. In the manuscript, we have inferred the real oxygen concentration of the 1111 phase is about 0.75 in the sample of $x=0.9$. Then we have 
$$\mathrm{ThCrAsN}_{0.1}\mathrm{O}_{0.9}~(\mathrm{nominal}) = m \mathrm{ThCrAsN}_{0.25}\mathrm{O}_{0.75} + n \mathrm{ThO}_{2} + \mathrm{other~phases}.$$
If we take N and O as the conservations, we have $0.25m = 0.1,0.75m + 2n = 0.9$. Then we get $m = 0.4, n = 0.3$. Although ThO$_2$ has a higher main peak than ThCrAsN$_{0.25}$O$_{0.75}$, the result that 1111 phase is slightly more than ThO$_2$ is still reasonable, because the diffraction peaks of ThO$_2$ are sharper.

\newpage

\subsection{X-ray diffraction for ThCrAsN$_{0.9}$O$_{0.1}$}

\begin{figure}[h]
	\centering
	\includegraphics[width=0.8\columnwidth]{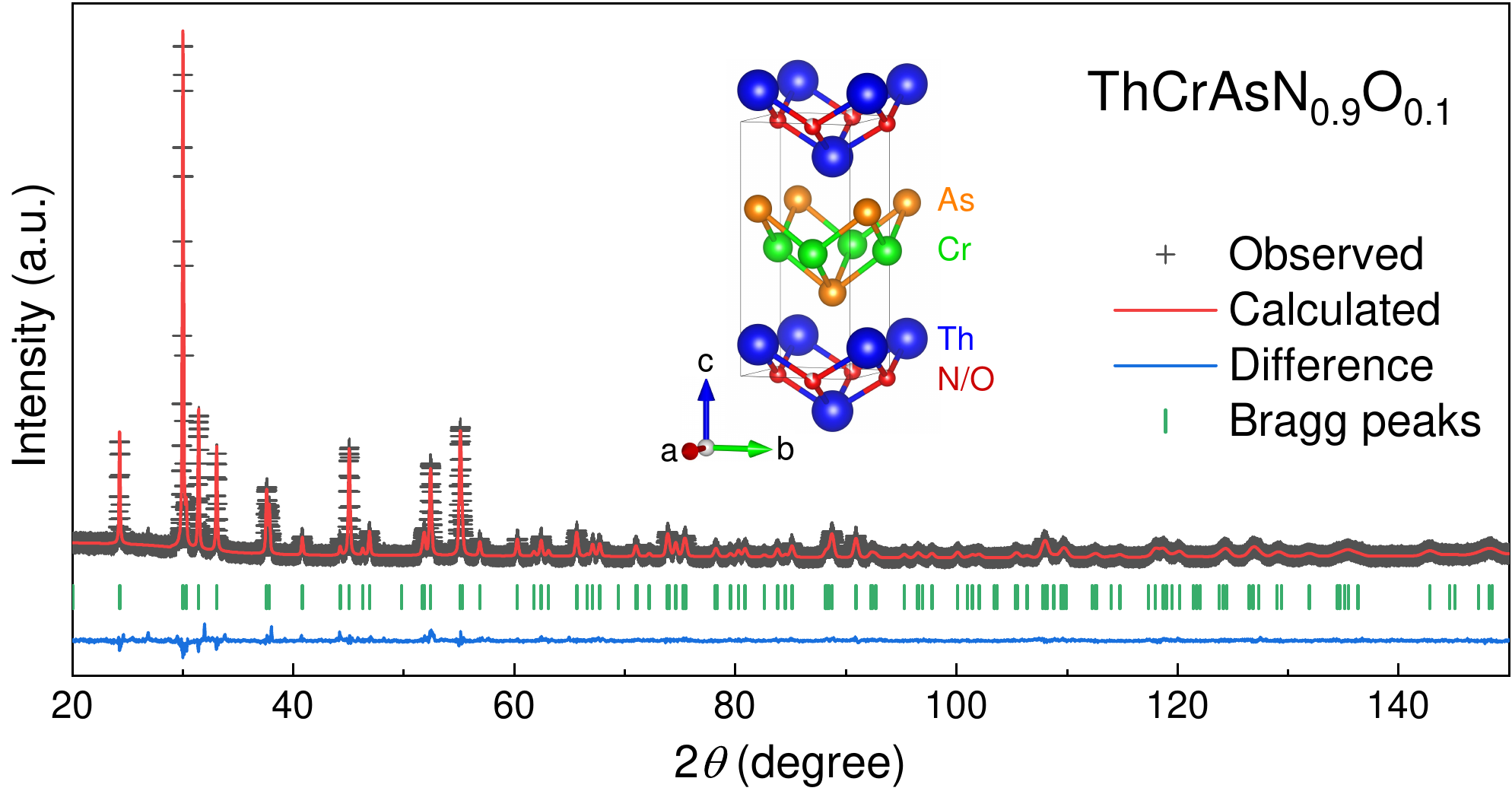}
	\caption{Powder X-ray diffraction data and the Rietveld analysis on sample of ThCrAsN$_{0.9}$O$_{0.1}$. Several tiny impurity peaks from Th$_3$As$_4$ could be seen in the pattern.}
	\label{fig:FigS1_XRD}
\end{figure}

\subsection{Crystallographic data of ThFeAsN$_{0.9}$O$_{0.1}$ at 300 K}
\begin{table}[h]
	\caption{\label{Tab1} Crystallographic data of ThCrAsN$_{0.9}$O$_{0.1}$ at 300 K obtained by the Rietveld refinement shown in Fig. \ref{fig:FigS1_XRD}. The cell parameters of ThCrAsN$_{0.9}$O$_{0.1}$ are: $a=4.0185(1)$ \AA, $c=8.8315(2)$ \AA.}
	\begin{ruledtabular}
		\begin{tabular}{ccccccc}	
			Atom & Wyckoff & occupancy & $x$ & $y$ & $z$ & $B_\mathrm{iso} (\mathrm{\AA}^{-2})$\\
			\hline
			Th & $2c$ & 1 & 0.25 & 0.25 & 0.1364(1) & 0.1\\
			Cr & $2a$ & 1 & 0.75  & 0.25 & 0.5 & 0.3\\
			As & $2c$ & 1 & 0.25 & 0.25 & 0.6704(3) & 0.3\\
			N & $2b$ & 0.9 & 0.75 & 0.25 & 0 & 1\\
			O & $2b$ & 0.1 & 0.75 & 0.25 & 0 & 1\\			
			
		\end{tabular}
	\end{ruledtabular}
\end{table}

\newpage

\subsection{Magnetic susceptibility for ThCrAsN$_{1-x}$O$_x$}
\begin{figure}[h]
	\centering
	\includegraphics[width=\columnwidth]{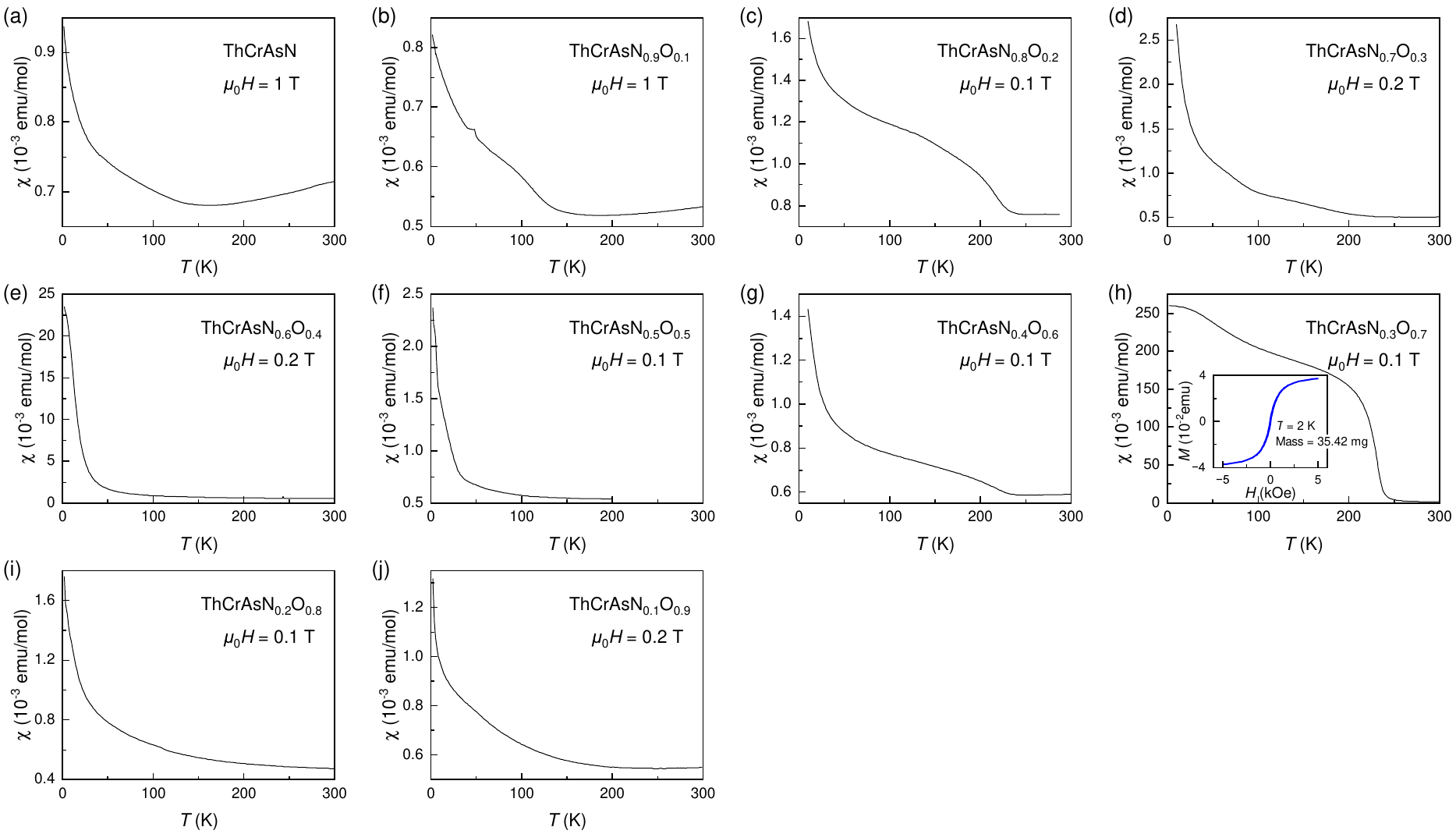}
	\caption{Temperature dependences of the magnetic susceptibility for ThCrAsN$_{1-x}$O$_x$ ($x\geq0.1$). The robust magnetic impurities exist in all samples, so there is no trends for the data with different oxygen concentration.}
	\label{fig:FigS2_Mag}
\end{figure}
\textbf{Discussion on the ferromagnetic impurities in Fig. S2}

The ferromagnetic (FM) behavior ($T_\mathrm{c}\sim$ 225 K) is observed in the panels (c)(g)(h), which does not come from ThCrAsN$_{1-x}$O$_{x}$, because it doesn’t follow obvious trends as a function of O substitution and the transition temperatures are all around 225 K. The FM impurities are not the other identified phases either, since they are nonmagnetic (ThO$_{2}$, Th$_{3}$As$_{4}$) or antiferromagnetic (Cr$_{2}$As). The possible candidate for the FM impurity is Cr$_3$As$_2$, which was reported to be FM with $T_\mathrm{c}$ around 225 K [J. Phys. Soc. Jpn. \textbf{15}, 2007 (1960).]. We can make a simple estimation with the $M(H)$ data of ThCrAsN$_{0.3}$O$_{0.7}$ [inset in panel (h)] to show the proportion of FM impurity in the sample. 

If we assume the magnetization of ThCrAsN$_{0.3}$O$_{0.7}$ sample at 5000 Oe (2 K) is all contributed by Cr$_3$As$_2$, which is 0.037 emu [see the inset of panel (h)], and the saturation magnetization of Cr$_3$As$_2$ is 21.0 emu/g [J. Phys. Soc. Jpn. \textbf{15}, 2007 (1960).]. Then the mass of Cr$_3$As$_2$ is calculated to be 1.77 mg, while the total mass of the sample is 35.42 mg. So the FM impurity in the sample of ThCrAsN$_{0.3}$O$_{0.7}$ is less than 5\%. Please note that the susceptibility of ThCrAsN$_{0.3}$O$_{0.7}$ is greater than the data of other samples by one or two orders of magnitude, so the FM impurity in other samples should be less than 1\%.


\subsection{Calculated magnetic energies and Cr spin moments}
\begin{table}[h]
	\caption{\label{Tab2} Magnetic energies and Cr spin moments for ThCrAsN and ThCrAsO with different spin configurations. Note that S-FM in the table denotes the magnetic structure with in-plane striped AFM order and out-of-plane FM coupling, i.e. the notation S in the Fig. 4 in the main text. S-AFM here denotes in-plane striped AFM order and out-of-plane AFM coupling, and FM denotes in-plane FM order and out-of-plane FM coupling. Other notations are the same as the ones in the main text.}
	\begin{ruledtabular}
		\begin{tabular}{ccccc}	
			& \multicolumn{2}{c}{ThCrAsN} & \multicolumn{2}{c}{ThCrAsO} \\			
			& $E_m$ (meV/f.u.) & moment ($\mu_\mathrm{B}$) & $E_m$ (meV/f.u.) & moment ($\mu_\mathrm{B}$) \\
			\hline
			NM & 0 & 0 & 0 & 0 \\
			A & $ -273.85 $ & 2.214 & $ -90.54 $ & 1.3 \\	
			C & $ -395.077 $ & 2.505 & $ -244.37 $ & 2.43 \\
			G & $ -397.14 $ & 2.507 & $ -246.60 $ & 2.43 \\			
			S-AFM & \multicolumn{2}{c}{Converge to NM} & \multicolumn{2}{c}{Converge to NM} \\
			S-FM & $ -193.17 $ & 2.451 & \multicolumn{2}{c}{Converge to NM} \\
			FM & $ -109.28 $ & 2.248 & $ -91.88 $ & 1.138 \\
			
		\end{tabular}
	\end{ruledtabular}
\end{table}

\newpage
\subsection{Band structures of ThCrAsN and ThCrAsO}
\begin{figure}[h]
	\centering
	\includegraphics[width=0.6\columnwidth]{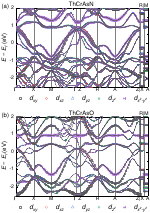}
	\caption{Electronic band structures near the Fermi energy of (a) ThCrAsN and (b) ThCrAsO. The contributions from different Cr $3d$ orbitals are distinguished by different symbols and colors. The symbols’ size represents the relative weight. The band structures for the two materials are overall similar, except the chemical potential of ThCrAsO is higher. Most of the bands are almost flat along the symmetry lines $\mathrm{\Gamma Z}$ and $\mathrm{MA}$, which are along the $c$ direction, consistent with the 2D layered structure of ThCrAsN$_{1-x}$O$_x$. }
	\label{fig:FigS3_band}
\end{figure}
